\documentclass[a4paper,preprint,superscriptaddress,showpacs,preprintnumbers,nofootinbib]{revtex4}
\usepackage{graphicx}
\usepackage{epstopdf}
\usepackage{dcolumn}
\usepackage{bm}
\usepackage{axodraw}

\newcommand{\nn}{\nonumber}

\newcommand{\be}{\begin{eqnarray}}
\newcommand{\ee}{\end{eqnarray}}
\catcode`\@=11
\def\lsim{\mathrel{\mathpalette\@versim<}}
\def\gsim{\mathrel{\mathpalette\@versim>}}
\def\@versim#1#2{\vcenter{\offinterlineskip
\ialign{$\m@th#1\hfil##\hfil$\crcr#2\crcr\sim\crcr } }}
\catcode`\@=12

\def\thefootnote{\fnsymbol{footnote}}

\begin{document}

\title{Multi-Component Dark Matter Systems and \\Their Observation Prospects}

\author{Mayumi \surname{Aoki}}
\email{mayumi@hep.s.kanazawa-u.ac.jp}
\affiliation{Institute for Theoretical Physics, Kanazawa University, Kanazawa 920-1192, Japan}
\affiliation{Max--Planck--Institut f\"ur Kernphysik, Saupfercheckweg 1, 69117 Heidelberg, Germany}

\author{Michael \surname{Duerr}}
\email{michael.duerr@mpi-hd.mpg.de}
\affiliation{Max--Planck--Institut f\"ur Kernphysik, Saupfercheckweg 1, 69117 Heidelberg, Germany}

\author{Jisuke \surname{Kubo}}
\email{jik@hep.s.kanazawa-u.ac.jp}
\affiliation{Institute for Theoretical Physics, Kanazawa University, Kanazawa 920-1192, Japan}

\author{Hiroshi \surname{Takano}}
\email{takano@hep.s.kanazawa-u.ac.jp}
\affiliation{Institute for Theoretical Physics, Kanazawa University, Kanazawa 920-1192, Japan}

\preprint{KANAZAWA-12-07}


\begin{abstract}
Conversions and semi-annihilations of dark matter (DM) particles 
in addition to the standard DM annihilations are considered in 
a three-component DM system. 
We find that the relic abundance of DM can be very sensitive to 
these non-standard DM annihilation processes, 
which has been  recently  found for 
two-component DM systems. 
To consider a concrete model of a three-component DM system, 
we extend the radiative seesaw model of Ma by adding 
a Majorana fermion $\chi$ and a real scalar boson $\phi$, 
to obtain a $Z_2\times Z'_2$ DM stabilizing symmetry, 
where we assume that the DM particles are the inert Higgs boson, 
$\chi$ and $\phi$.
It is shown how the allowed parameter space, obtained previously 
in the absence of $\chi$ and $\phi$, changes.
The semi-annihilation process in this model produces  monochromatic neutrinos.
The observation rate of 
 these monochromatic neutrinos from the Sun at IceCube is estimated.
Observations of high energy monochromatic neutrinos from the Sun may 
indicate a multi-component DM system.

 \end{abstract}
\setcounter{footnote}{0}
\def\thefootnote{\arabic{footnote}}
\maketitle

\section{Introduction}

Recent astrophysical observations 
\cite{Riess:1998cb,Abazajian:2008wr,Komatsu:2010fb} 
have made it clear  that most of the energy 
of the Universe consists of dark energy and cold dark matter (DM), 
and their portions are very well fixed by these observations. 
While the origin of dark energy might be the cosmological constant 
of Einstein, the origin of cold DM cannot be found 
within the framework of the standard model (SM) 
of elementary particles.
Moreover, we do not know very much about the detailed features of 
DM at present, even if the origin of  DM should be elementary particles.
Currently, many experiments are undertaken or planned, 
and it is widely believed that the existence of DM will be 
independently confirmed in the near future (see, for instance, Refs.~\cite{Jungman:1995df,Bertone:2004pz,Cirelli:2012tf}).

A particle DM candidate can be made stable by an unbroken symmetry.
The simplest possibility of such a symmetry is a parity, $Z_2$.
Whatever the origin of the $Z_2$ is, the lightest $Z_2$-odd particle
can be a DM candidate if it is a neutral, weakly interacting
and massive particle (WIMP)  (see Ref.~\cite{Bertone:2004pz} for a review).
There are a variety of origins of the $Z_2$.
$R$ parity in the minimal
supersymmetric standard model (MSSM), which is introduced 
to forbid fast proton decay, is a well-known example
(see Ref.~\cite{Jungman:1995df} for a review).
In this paper, we consider a universe
consisting of stable multi-DM particles
 \cite{Berezhiani:1990sy}--\cite{Aoki:2011he}.
A multi-component DM system  can be realized if the
 DM stabilizing symmetry is larger than $Z_2$:
$Z_N~(N \geq 4)$ or
a product of two or more $Z_2$'s
can yield a multi-component DM system.\footnote{$Z_3$ allows only one-component 
DM systems. Refs.~\cite{Ma:2007gq,Agashe:2010gt} discuss models with  $Z_3$.} 
In a supersymmetric extension
of the radiative seesaw model of Ref.~\cite{Ma:2006km}, for instance, 
a $Z_2 \times Z'_2$ symmetry appears, 
providing various concrete models of multi-component 
DM systems \cite{Ma:2006uv}--\cite{Aoki:2011he}.

In a multi-component DM system, there can be  various  DM annihilation 
processes that are different from the standard DM annihilation
process \cite{Griest:1988ma}--\cite{Drees:1992am}, 
$\mbox{DM}~ \mbox{DM}\to X X$, where $X$ is
a generic SM particle in thermal equilibrium. 
Even in one-component DM systems, the non-standard annihilation
process, the co-annihilation of DM and a nearly degenerate 
unstable particle \cite{Griest:1990kh}, can  play a crucial  role in the
 MSSM \cite{Ellis:1998kh}.
The importance of non-standard annihilation
processes such as DM conversion
 \cite{D'Eramo:2010ep,Belanger:2011ww,Belanger:2012vp} and 
semi-annihilation of DM 
 \cite{D'Eramo:2010ep,Belanger:2012vp} in  two-component
DM systems 
for the temperature evolution of the number density of DM has been recently reported.

If  $(Z_2)^\ell$ is unbroken, there can exist at least $K=\ell$ stable  DM particles.
In a   kinematically fortunate situation,  $2^\ell-1$ stable DM particles
can exist; for $\ell=2$ there can be maximally $K=3$ stable DM particles.
Any one-component DM model can easily be 
extended to a multi-component DM system.
The allowed  parameter space of a one-component DM model
can considerably change,
as has been recently found in Ref.~\cite{Aoki:2011he} (see also Ref.~\cite{Cao:2007fy}),
 even using  a crude approximation of a DM conversion  process
 in a supersymmetric extension of the radiative seesaw model.

In Sec.~II, after outlining a derivation of 
 the  coupled Boltzmann equations
 that are appropriate for our purpose,
we consider fictive two- and three-component DM systems
 and analyze the effects of non-standard annihilation processes of DM. 
In Sec.~III we extend the radiative seesaw model of Ref.~\cite{Ma:2006km}
by adding an extra  Majorana fermion $\chi$ and 
an extra real scalar boson $\phi $, so as to obtain $Z_2\times Z'_2$
as a DM stabilizing symmetry.
Apart from the presence  of $\phi $, the Higgs sector is identical to
that of Refs.~\cite{Barbieri:2006dq,LopezHonorez:2006gr,Dolle:2009fn}.
This model shows how the allowed parameter space, which is
obtained in Refs.~\cite{Barbieri:2006dq,LopezHonorez:2006gr,Dolle:2009fn}
under the assumption that the lightest inert Higgs boson
is DM,  can change.
Indirect  detection of DM---in particular, of neutrinos from the annihilation of the 
captured DM in the Sun \cite{Silk:1985ax}--\cite{Hooper:2002gs} ---is also discussed.
We solve the coupled evolution equations 
of the DM numbers in the Sun,
which describe approaching
 equilibrium between the capture  and annihilation 
 (including conversion and semi-annihilation) rates of DM,
 and estimate the observation rates of neutrinos.
Due to semi-annihilations of DM,
 monochromatic neutrinos are radiated from the Sun.
 Our conclusions are given in Sec.~IV.

\section{The Boltzmann equation and two- and three-component DM systems}
\subsection{The Boltzmann equation}
\begin{figure}[htb]
\begin{center}
\SetWidth{0.7}
  \includegraphics[clip]{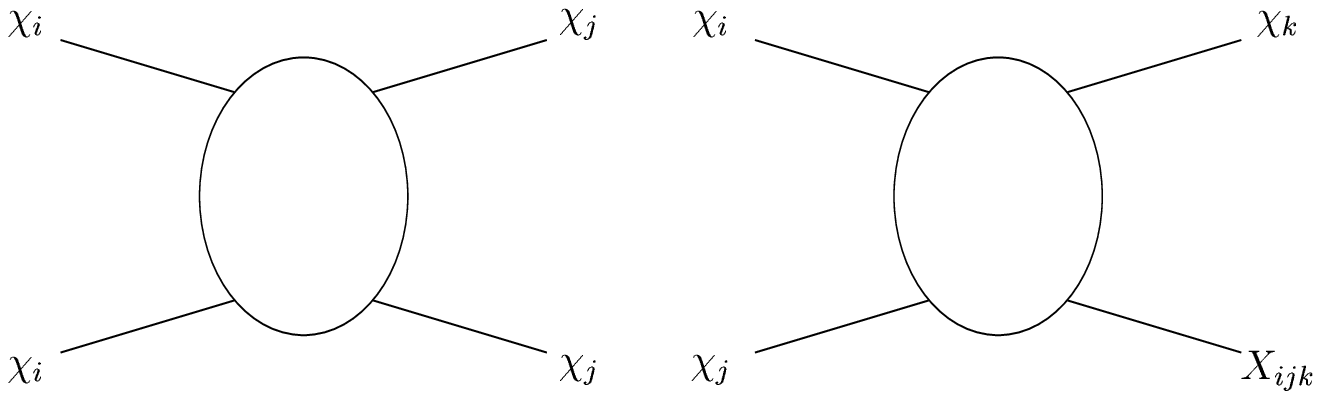}
\end{center}
\caption[]{\label{NS2} Dark matter conversion (left) and semi-annihilation (right).
}
\end{figure}

Here we would like to outline  a derivation of 
 the  Boltzmann equation
 that we are going to apply in the 
 next section. We will do it for  completeness,
 although the following discussion partially
 parallels that of Ref.~\cite{D'Eramo:2010ep} (see also Ref.~\cite{Belanger:2011ww}).
We start by assuming the existence of $K$ stable DM particles $\chi_i$
with mass $m_i$. None of the DM particles have the same quantum number
with respect to the DM stabilizing symmetry. 
All the other particles are supposed to be in thermal
equilibrium. Then we restrict ourselves to   three types of processes
which enter the Boltzmann equation:
\be
& &\chi_i~\chi_i \leftrightarrow X_i~X'_i~,
\label{p0}\\
 & &  \chi_i~\chi_i  \leftrightarrow \chi_j~\chi_j~
 \mbox{(DM conversion)}~,
 \label{p1}\\
& & \chi_i~\chi_j \leftrightarrow \chi_k~X_{ijk}~
 \mbox{(DM semi-annihilation)}~,
\label{p2}
\ee
where the extension to include coannihilations and annihilation processes like 
 $\chi_i+\chi_j  \leftrightarrow \chi_k+\chi_l$ is straight forward. 
See Fig.~\ref{NS2} for a depiction of DM conversion and DM semi-annihilation. 

We  denote  the phase space density of $\chi_i$ by $f_i (E_i,t)$ and its number density
by $n_i(t)=(g/(2\pi)^3)\int  d^3 p_i f_i(E_i,t)$, where $g$ stands for the internal
degrees of freedom. 
Then the density $n_i$ satisfies the Boltzmann equation
(see, e.g., Ref.~\cite{Kolb:1990vq}), which we will  not spell out here.
Instead, we make the replacement  
\be
t &=& 0.301 g_*^{-1/2} M_{\rm PL}T^{-2}~,
\label{approx1}
\ee
during the radiation-dominated epoch, 
where $t$ is the time of the comoving frame, 
$g_*$ is the total number of effective
degrees of freedom, and $ T$ and $M_{\rm PL}$ are the temperature and  the Planck mass,
respectively. Further, we use the approximation
\be
\frac{f_i (E_i,t)}{\bar{f}_i (E_i,t)} &\simeq & \frac{n_i (t)}{\bar{n}_i (t)}~,
\ee
where $\bar{f}_i (E_i,t)\simeq \exp (-E_i/T)$ and 
$\bar{n}_i=(g/(2\pi)^3)\int  d^3 p_i \bar{f}_i(E_i,t)$ are 
the values in equilibrium, and we ignore the chemical potential.
Then  the collision terms in the Boltzmann equation can be written as
\be
& & -(\mbox{PSI})| M(ii;X_i X'_i)|^2 \frac{\bar{f}_i\bar{f}_i}{\bar{n}_i\bar{n}_i}
\left(  n_i n _i-\bar{n}_{i}\bar{n}_{i}
\right)\nn\\
& &-\sum_{i >j}(\mbox{PSI})~| M(ii;jj)|^2 
\frac{\bar{f}_i\bar{f}_i}{\bar{n}_i\bar{n}_i}
\left(  n_i n _i-\frac{n_j n_j}
{\bar{n}_{j}\bar{n}_{j}} \bar{n}_{i}\bar{n}_{i}
\right)\nn\\
& &+\sum_{j >i}(\mbox{PSI})~| M(jj;ii)|^2 
\frac{\bar{f}_j\bar{f}_j}{\bar{n}_j\bar{n}_j}
\left(  n_j n _j-\frac{n_i n_i}
{\bar{n}_{i}\bar{n}_{i}} \bar{n}_{j}\bar{n}_{j}
\right)\nn\\
& &-\sum_{j,k}(\mbox{PSI})~| M(ij;k X_{ijk})|^2 
\frac{\bar{f}_i\bar{f}_j}{\bar{n}_i\bar{n}_j}
\left(  n_i n _j-\frac{n_k}
{\bar{n}_{k}} \bar{n}_{i}\bar{n}_{j}
\right)\nn\\
& &+\sum_{j,k}(\mbox{PSI})~| M(jk;i X_{jki})|^2 
\frac{\bar{f}_j\bar{f}_k}{\bar{n}_j\bar{n}_k}
\left(  n_j n _k-\frac{n_i}
{\bar{n}_{i}} \bar{n}_{j}\bar{n}_{k}
\right)~,
\label{collision}
\ee
where PSI stands for "phase space integral of $(2\pi)^4
\delta^4 ({\mbox{momenta}})\times$",
$M$ is the matrix element of the corresponding process, and
we  have assumed that
\be
m_i &\geq & m_j~\mbox{for}~i >j~~\mbox{and}
~~m_{X_i},  m_{X'_i}, m_{X_{ijk}} << m_l
~\mbox{for all}~i,j,k,l~.
\ee
Using the notion of the thermally averaged cross section,
\be
<\sigma (ii;X_i X'_i) v> &=&\frac{1}{\bar{n}_{i}\bar{n}_{i}}
\mbox{PSI} | M(ii;X_iX'_i)|^2 \bar{f}_{i}\bar{f}_{i}~,
\ee
and the dimensionless inverse temperature $x=\mu/T$, we obtain for the number per comoving volume, $Y_i=n_i/s$:
\be
& &\frac{d Y_i}{d x}=-0.264~ g_*^{1/2} \left[\frac{\mu M_{\rm PL}}{x^2} \right]
\Big\{~
<\!\sigma (ii;X_i X'_i) v\!>\left(  Y_i Y _i-\bar{Y}_{i}\bar{Y}_{i}\right)
\nn\\
& &\left. +
\sum_{i >j }<\!\sigma (ii;jj)v\!>\!\!\left(  Y_i Y _i-\frac{Y_j Y_j}
{\bar{Y}_{j}\bar{Y}_{j}} \bar{Y}_{i}\bar{Y}_{i}
\right)
\!-\!\sum_{j >i}<\!\sigma (jj;ii)v\!>\!\!
\left(  Y_j Y _j-\frac{Y_i Y_i}
{\bar{Y}_{i}\bar{Y}_{i}} \bar{Y}_{j}\bar{Y}_{j}
\right)~\right.~\nn\\
& &\!+\!\sum_{j,k}<\!\sigma (ij;k X_{ijk})v\!>\!\!
\left(  Y_i Y _j-\frac{Y_k}
{\bar{Y}_{k}} \bar{Y}_{i}\bar{Y}_{j}
\right)-\!\sum_{j,k}<\!\sigma (jk;i X_{jki})v\!>\!\!
\left(  Y_j Y _k-\frac{Y_i}
{\bar{Y}_{i}} \bar{Y}_{j}\bar{Y}_{k}
\right)\Big\}~,
\label{boltz}
\ee
where $1/\mu=(\sum_i m_i^{-1})$ is the reduced mass of the system.
To arrive at Eq.~(\ref{boltz}), we have used $s=(2 \pi^2/45 ) g_* T^3~,~
H=1.66\times g_*^{1/2} T^2/M_{\rm PL}$, where 
$s$ is the entropy density and $H$ is the Hubble constant.

We will integrate this system of coupled non-linear
differential equations numerically.
Before we apply  the Boltzmann equation [Eq.~(\ref{boltz})]
to a concrete DM model, we discuss below
the cases of $K=2$ and $3$, simply assuming
fictitious values of the   thermally averaged cross sections
and DM masses $m_i$.

\subsection{Two-component DM system}
Before we come to one of our main interests, a three-component DM system, we first  consider
the $K=2$ case with a $Z_2 \times Z'_2$ symmetry.
In this case, there are three different 
   thermally averaged cross sections. 
 No semi-annihilation [Eq.~(\ref{p2})] is  allowed  due 
 to $Z_2 \times Z'_2$.\footnote{In Refs.~\cite{D'Eramo:2010ep,Belanger:2012vp}, the $Z_4$ case
is discussed in detail. In this case  there exist two independent DM particles,
because due to CP invariance, the anti-particle is  not an independent degree of 
freedom in the Boltzmann equation. Semi-annihilation is allowed in this case.}
 We further assume that there are only 
 $s$-wave contributions to $<\! \sigma v \!>$
 and that $X_i ~(i=1,2)$ are massless while $m_1\geq m_2 $:
 \be
 <\!\sigma (11;X_1 X'_1) v\!>
 &=& \sigma_{0,1}\times 10^{-9} ~\mbox{GeV}^{-2}~,~
 <\!\sigma (22;X_2 X'_2) v\!>=\sigma_{0,2}\times 10^{-9}~\mbox{GeV}^{-2}~,\nn\\
  <\!\sigma (11;22 ) v\!> &=&
\sigma_{0,12}\times 10^{-9}~\mbox{GeV}^{-2}~.
    \label{input1}
   \ee
  Equation~(\ref{boltz}) then becomes
  \be
& &\frac{d Y_1}{dx}=
-0.264~ g_*^{1/2} \left[\frac{\mu M_{\rm PL}}{x^2} \right]
\left\{<\!\sigma (11;X_1 X'_1) v\!>
\left(  Y_1 Y _1-\bar{Y}_{1}\bar{Y}_{1}\right)\right.\nn\\
& &\left.
+<\!\sigma (11;22)v\!>\!\!\left(  Y_1 Y _1-\frac{Y_2 Y_2}
{\bar{Y}_{2}\bar{Y}_{2}} \bar{Y}_{1}\bar{Y}_{1}
\right)\right\}~,\label{boltz21}\\
& &\frac{d Y_2}{dx}=-0.264~ g_*^{1/2} \left[\frac{\mu M_{\rm PL}}{x^2} \right]
\left\{ <\!\sigma (22;X_2X'_2) v\!>\left(  Y_2 Y _2-\bar{Y}_{2}\bar{Y}_{2}\right)
\right.\nn\\
& &\left.
-<\!\sigma (11;22)v\!>\!\!\left(  Y_1 Y _1-\frac{Y_2 Y_2}
{\bar{Y}_{2}\bar{Y}_{2}} \bar{Y}_{1}\bar{Y}_{1}
\right)\right\}~.
\label{boltz22}
\ee

\begin{figure}
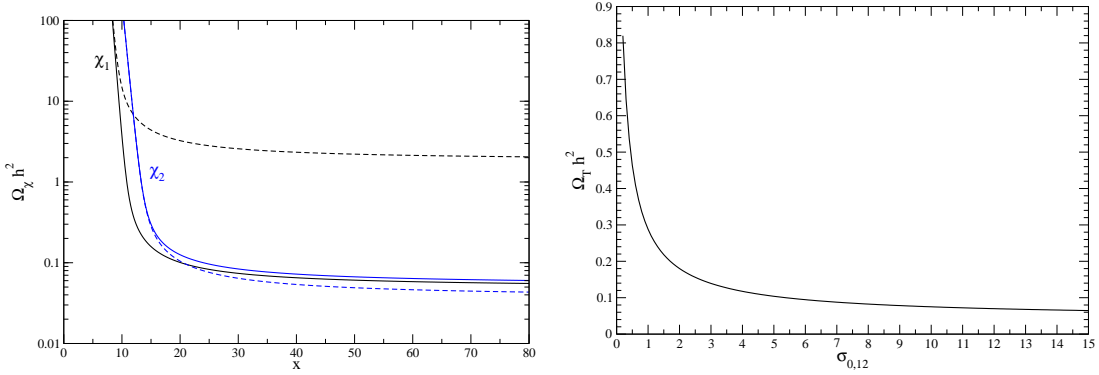

  \includegraphics[width=7cm]{omega12.eps}
 \hspace{0.1cm}   \includegraphics[width=7cm]{omegaT-2.eps}
\caption{\label{omega12}\footnotesize
Left: The relic abundance $\Omega_{\chi_1}h^2(x)$ (black curves) and 
$\Omega_{\chi_2} h^2(x)$ (blue curves) as a function of $x=\mu/T=[(m_1^{-1}+m_2^{-1})T]^{-1}$
 with $\sigma_{0,1}=0.1~,~
\sigma_{0,2}=6~,~
\sigma_{0,12}=4.4$ (solid curves) or 0 (dashed curves)~,~
$m_1=200$ GeV, $m_2=160$ GeV and $g_*=90$.
Right: The total relic abundance $\Omega_Th^2$ as a function of 
 $\sigma_{0,12}$ which parametrizes the size of the conversion 
 $\chi_1 \chi_1 \to \chi_2\chi_2$.
}
\end{figure}

We consider the case in which  the size of 
 the DM conversion  and  the standard annihilation are 
of  similar order (see also Ref.~\cite{Belanger:2011ww}).
 In Fig.~\ref{omega12} (left), we show the evolution of the fraction of critical densities,
 $\Omega_{\chi_1}h^2(x)$ (black curves) and $\Omega_{\chi_2} h^2(x)$ (blue curves), contributed
 by $\chi_1$ and $\chi_2$, respectively,
 where we have used 
$\sigma_{0,1}=0.1~,~
\sigma_{0,2}=6~,~
\sigma_{0,12}=4.4~\mbox{(solid curves)}$ or 
$0~\mbox{(dashed curves)}$,
$m_1=200$ GeV, $m_2=160$ GeV, $g_*=90$, and
$x=\mu/T=[(m_1^{-1}+m_2^{-1})T]^{-1}$.
As we see from Fig.~\ref{omega12} (left),
at $\sigma_{0,12}=0$ (i.e., no DM conversion, Eq.~(\ref{p1})),
the density of $\chi_1$  decouples from the equilibrium value
for smaller $x$ than the density of $\chi_2$ does.
This is because we have chosen a small value for 
$\sigma_{0,1}$ and a large value for $\sigma_{0,2}$.
At $\sigma_{0,12}=0$, $\Omega_{\chi_1}h^2 \approx 1.99$, while
  $\Omega_{\chi_2}h^2\approx 0.04$.
With increasing value of $\sigma_{0,12}$ (which parametrizes the size of
the DM conversion, Eq.~(\ref{p1})),
 $\Omega_{\chi_1}h^2$ decreases, while 
 $\Omega_{\chi_2}h^2$ increases.
 Around $\sigma_{0,12}=3.9$, this order changes, i.e.,
  $\Omega_{\chi_1} < \Omega_{\chi_2}$. 
  At  $\sigma_{0,12}=4.4$, we obtain 
 the total relic abundance 
 $\Omega_T h^2=\Omega_{\chi_1} h^2+\Omega_{\chi_2} h^2=0.112$,
  in accord with the WMAP observation $\Omega_T h^2=0.1126\pm 0.0036$
  \cite{Komatsu:2010fb}.
 In Fig.~\ref{omega12} (right), we plot $\Omega_T h^2$
 as a function of $\sigma_{0,12}$. We see
 that the DM conversion process plays an important role, as has also been found
 in Refs.~\cite{SungCheon:2008ts,D'Eramo:2010ep,Belanger:2011ww,Belanger:2012vp}.

\subsection{Three-component DM system}
As we have noticed before, the $K=3$ case is possible even for a
$Z_2\times Z'_2$ symmetry if the decay of $\chi_i$ is kinematically
forbidden. In this case, there are nine different  thermally-averaged cross sections,
if we assume that $m_1 \geq m_2 \geq m_3$ and $m_2+m_3  > m_1$:
 \be
 <\!\sigma (ii;X_i X'_i) v\!>
 &=& \sigma_{0,i}\times 10^{-9} ~\mbox{GeV}^{-2}~,~
  <\!\sigma (11;22 ) v\!> = \sigma_{0,12}\times 10^{-9}~\mbox{GeV}^{-2}~,\nn\\
 <\!\sigma (11;33) v\!>
 &=& \sigma_{0,13}\times 10^{-9} ~\mbox{GeV}^{-2}~,~
  <\!\sigma (22;33 ) v\!> = \sigma_{0,23}\times 10^{-9}~\mbox{GeV}^{-2}~,\nn\\
 <\!\sigma (12;3X_{123}) v\!>
 &=& \sigma_{0,123}\times 10^{-9} ~\mbox{GeV}^{-2}~,~
  <\!\sigma (23;1X_{231} ) v\!> = \sigma_{0,231}\times 10^{-9}~\mbox{GeV}^{-2}~,
  \nn\\
  <\!\sigma (31;2X_{312}) v\!>
 &=& \sigma_{0,312}\times 10^{-9} ~\mbox{GeV}^{-2}~.
 \label{sigma3}
   \ee

 Equation (\ref{boltz}) then becomes
  \be
& &\frac{d Y_1}{dx}=
-0.264~ g_*^{1/2} \left[\frac{\mu M_{\rm PL}}{x^2} \right]
\left\{<\!\sigma (11;X_1 X'_1) v\!>
\left(  Y_1 Y _1-\bar{Y}_{1}\bar{Y}_{1}\right)\right.\nn\\
& &
+<\!\sigma (11;22)v\!>\!\!\left(  Y_1 Y _1-\frac{Y_2 Y_2}
{\bar{Y}_{2}\bar{Y}_{2}} \bar{Y}_{1}\bar{Y}_{1}
\right)
+<\!\sigma (11;33)v\!>\!\!\left(  Y_1 Y _1-\frac{Y_3 Y_3}
{\bar{Y}_{3}\bar{Y}_{3}} \bar{Y}_{1}\bar{Y}_{1}
\right)~\nn\\
& &
+<\!\sigma (12;3X_{123})v\!>\!\!\left(  Y_1 Y _2-\frac{Y_3}
{\bar{Y}_{3}} \bar{Y}_{1}\bar{Y}_{2}
\right)+<\!\sigma (31;2X_{312})v\!>\!\!\left(  Y_1 Y _3-\frac{Y_2}
{\bar{Y}_{2}} \bar{Y}_{1}\bar{Y}_{3}
\right)\nn\\
& &\left.-<\!\sigma (23;1X_{231})v\!>\!\!\left(  Y_3 Y _2-\frac{Y_1}
{\bar{Y}_{1}} \bar{Y}_{3}\bar{Y}_{2}
\right)\right\}~,\\
& &\frac{d Y_2}{dx}=
-0.264~ g_*^{1/2} \left[\frac{\mu M_{\rm PL}}{x^2} \right]
\left\{<\!\sigma (22;X_2 X'_2) v\!>
\left(  Y_2 Y _2-\bar{Y}_{2}\bar{Y}_{2}\right)\right.\nn\\
& &
+<\!\sigma (22;33)v\!>\!\!\left(  Y_2 Y _2-\frac{Y_3 Y_3}
{\bar{Y}_{3}\bar{Y}_{3}} \bar{Y}_{2}\bar{Y}_{2}\right)
+<\!\sigma (23;1X_{231})v\!>\!\!\left(  Y_2 Y _3-\frac{Y_1}
{\bar{Y}_{1}} \bar{Y}_{2}\bar{Y}_{3}
\right)\nn\\
& &\left.
+<\!\sigma (12;3X_{123})v\!>\!\!\left(  Y_1 Y _2-\frac{Y_3}
{\bar{Y}_{3}} \bar{Y}_{1}\bar{Y}_{2}
\right)-<\!\sigma (31;2X_{312})v\!>\!\!\left(  Y_1 Y _3-\frac{Y_2}
{\bar{Y}_{2}} \bar{Y}_{1}\bar{Y}_{3}
\right)\right.\nn\\
& &\left.-<\!\sigma (11;22)v\!>\!\!\left(  Y_1 Y _1-\frac{Y_2 Y_2}
{\bar{Y}_{2}\bar{Y}_{2}} \bar{Y}_{1}\bar{Y}_{1}
\right)\right\}~,\\
& &\frac{d Y_3}{dx}=
-0.264~ g_*^{1/2} \left[\frac{\mu M_{\rm PL}}{x^2} \right]
\left\{<\!\sigma (33;X_3 X'_3) v\!>
\left(  Y_3 Y _3-\bar{Y}_{3}\bar{Y}_{3}\right)\right.\nn\\
& &
+<\!\sigma (23;1X_{231})v\!>\!\!\left(  Y_2 Y _3-\frac{Y_1}
{\bar{Y}_{1}} \bar{Y}_{2}\bar{Y}_{3}
\right)
+<\!\sigma (31;2X_{312})v\!>\!\!\left(  Y_1 Y _3-\frac{Y_2}
{\bar{Y}_{2}} \bar{Y}_{1}\bar{Y}_{3}
\right)\nn\\
& &\left.-<\!\sigma (12;3X_{123})v\!>\!\!\left(  Y_1 Y _2-\frac{Y_3}
{\bar{Y}_{3}} \bar{Y}_{1}\bar{Y}_{2}
\right)-<\!\sigma (11;33)v\!>\!\!\left(  Y_1 Y _1-\frac{Y_3 Y_3}
{\bar{Y}_{3}\bar{Y}_{3}} \bar{Y}_{1}\bar{Y}_{1}
\right)\right.\nn\\~
& &\left.-<\!\sigma (22;33)v\!>\!\!\left(  Y_2 Y _2-\frac{Y_3 Y_3}
{\bar{Y}_{3}\bar{Y}_{3}} \bar{Y}_{2}\bar{Y}_{2}
\right)~\right\}~,
\label{boltz3}
\ee
where $1/\mu=1/m_1+1/m_2+1/m_3$.

\begin{figure}
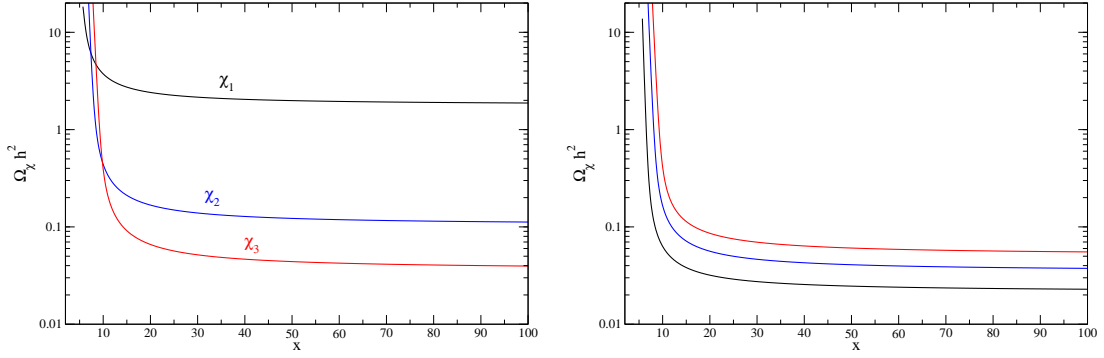

  \includegraphics[width=7cm]{omega123-0.eps}
 \hspace{0.1cm}     \includegraphics[width=7cm]{omega123-52.eps}
\caption{\label{omega123-0}\footnotesize
The relic abundance $\Omega_{\chi_1}h^2(x)$ (black curve),
$\Omega_{\chi_2} h^2(x)$ (blue curve) and $\Omega_{\chi_3}h^2(x)$ (red curve)
as a function of $x=\mu/T=[(m_1^{-1}+m_2^{-1}+m_3^{-1})T]^{-1}$, where
 the input parameters are given in Eq.~(\ref{input2}).
Left: Without the non-standard annihilation processes  [Eqs.~(\ref{p1}) and (\ref{p2})]. 
Right: $\sigma_{0,12}=\sigma_{0,13}=\sigma_{0,23}=5.2$, while
$\sigma_{0,123}= \sigma_{0,312}=\sigma_{0,231}=0$, to see the effects
of $ \chi_i \chi_i \leftrightarrow \chi_j \chi_j$ type processes [Eq.~(\ref{p2})].}
\end{figure}

As a representative example, we consider 
the following set of input values of the parameters:
\be
m_1 &= & 200 ~\mbox{GeV}~,~
m_2 = 160 ~\mbox{GeV}~,~m_3 = 140 ~\mbox{GeV}~,\nn\\
\sigma_{0,1}&=& 0.1~,~\sigma_{0,2}=2~,~
\sigma_{0,3}=6~.
\label{input2}
\ee
First, we show the evolution of $\Omega_{\chi_i} h^2 (x)$
in Fig.~\ref{omega123-0} (left) for 
$\sigma_{0,12}=\sigma_{0,13}=\sigma_{0,23}=\sigma_{0,123}=
\sigma_{0,312}=\sigma_{0,231}=0$, which corresponds to
the situation without the non-standard annihilation processes.
Since $m_1 >m_2,m_3$, and the cross section  $\sigma (11;X_1 X_1)$
is small in this example, the relic abundance of $\chi_1$ is large
compared with that of $\chi_2$ and $\chi_3$.
This changes if we switch on the non-standard annihilation processes.
This is shown in Fig.~\ref{omega123-0} (right),
where we have used
$\sigma_{0,12}=\sigma_{0,13}=\sigma_{0,23}=5.2$, while
$\sigma_{0,123}=
\sigma_{0,312}=\sigma_{0,231}=0$, to show the effects
of $ \chi_i \chi_i \leftrightarrow \chi_j \chi_j$-type processes
(DM conversion).
As expected, the relic abundances of 
$\chi_1$ and $\chi_2$ decrease and drop below
$0.1$, while that of $\chi_3$ does not change very much.

\begin{figure}
  \includegraphics[width=8cm]{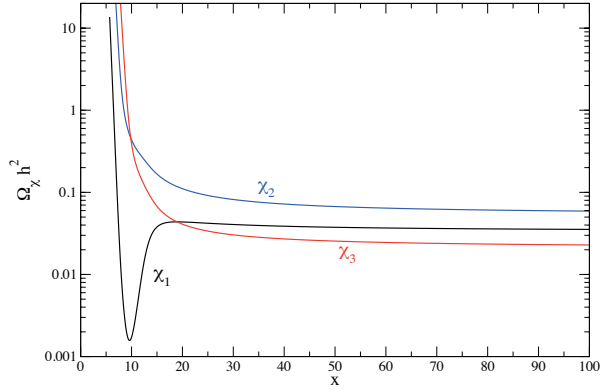}
\caption{\label{omega123-51}\footnotesize
The relic abundance $\Omega_{\chi_1}h^2(x)$ (black curve),
$\Omega_{\chi_2} h^2(x)$ (blue curve) and $\Omega_{\chi_3}h^2(x)$ (red curve)
as a function of $x$ with
$\sigma_{0,12}=\sigma_{0,13}=\sigma_{0,23}=0$, while
$\sigma_{0,123}=
\sigma_{0,312}=\sigma_{0,231}=5.1$, to show the effects
of $ \chi_i \chi_j \leftrightarrow \chi_k X_{ijk}$ type processes [Eq.~(\ref{p2})].}
\end{figure}

\begin{figure}
  \includegraphics[width=7cm]{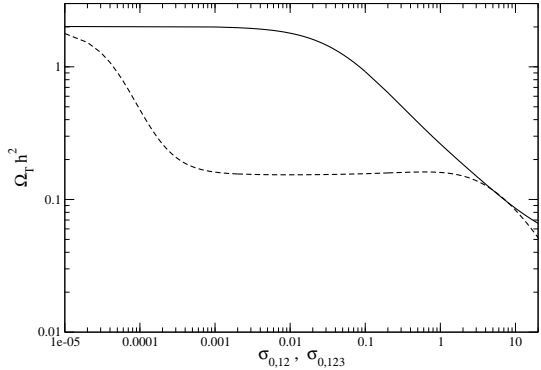}
\caption{\label{omega3T}\footnotesize
 The total relic abundance $\Omega_Th^2$ as a function of 
 $\sigma_{0,12}$ (solid curve)  and $\sigma_{0,123}$ (dashed curve).
 Except $\sigma_{0,12}$ (DM conversion) and  
 $\sigma_{0,123}$ (semi-annihilation), 
 the input parameters are given in Eq.~(\ref{input2}).
}
\end{figure}

Figure \ref{omega123-51} shows the evolution of $\Omega_{\chi_i} h^2 (x)$
for 
$\sigma_{0,12}=\sigma_{0,13}=\sigma_{0,23}=0$, while
$\sigma_{0,123}=
\sigma_{0,312}=\sigma_{0,231}=5.1$, to show the effects
of $ \chi_i \chi_j \leftrightarrow \chi_k X_{ijk}$-type processes
(semi-annihilation).
It is interesting to observe that
the order of the relic abundances changes, and
$\Omega_{\chi_1} h^2(x)$ first decreases as usual,
but then starts to increase towards the freeze-out value.
So, the effects of $\chi_i \chi_i \leftrightarrow \chi_j \chi_j$-type
and $ \chi_i \chi_j \leftrightarrow \chi_k X_{ijk}$-type processes
are different.
In the examples above, $\sigma_{0,ij}$ and $\sigma_{0,ijk}$ are chosen such that 
the total abundance $\Omega_T h^2$ becomes about the realistic value $0.112$.
In Fig.~\ref{omega3T}, we show the total abundance 
$\Omega_T h^2$ as a function of 
$\sigma_{0,ij}$ (solid curve) and $\sigma_{0,ijk}$ (dashed curve), where
$\sigma_{0,ij}$ parameterizes the size of the DM conversion [Eq.~(\ref{p1})]
and $\sigma_{0,ijk}$  parameterizes  the size of the semi-annihilation [Eq.~(\ref{p2})].
As we can see from Fig.~\ref{omega3T}, only for small values of $\sigma_{0,ij}$ and $\sigma_{0,ijk}$ are
the effects on $\Omega_T h^2$ different.

Note that the dark matter conversion process [Eq.~(\ref{p1})] is dark-matter-number conserving,
while 
the semi-annihilation process [Eq.~(\ref{p2})]  is not.
Next, we would like to consider an extreme case where
 only semi-annihilations  are present, and as before we assume that $m_1 \geq m_2 \geq m_3$
and $m_2+m_3 > m_1$.
In Table I, we show various  examples of the individual relic abundances
 with $m_1$ fixed at $1000$ GeV, where
we have assumed that the value of $\sigma_{0,ijk}$ is the same independent of $i,j$ and $k$.
These values are chosen such that  the total  relic abundance
is consistent with $\Omega_T h^2=0.1126\pm 0.0036$.
\begin{table}
\caption{\footnotesize{The relic abundances for the  symmetric case
of $\sigma_{0,ijk}$; i.e.,
$\sigma_{0,123}=\sigma_{0,231}=\sigma_{0,312}$,  with 
$m_1=1000 $ GeV.}
}
\begin{center}
\begin{tabular}{|c|c||c|c|c|c|} 
\hline
 $m_2$& $m_3$ &  $\sigma_{0,123}$  & $\Omega_{\chi_1} h^2$
& $\Omega_{\chi_2} h^2$& $\Omega_{\chi_3} h^2$\\ \hline
$720$& $700$  & $12.6$ & $0.0433$
& $0.0319$& $0.0372$\\ \hline
$940$  &  $700$ & $417.0$  & $0.0007$
& $0.0007$& $0.1109$\\ \hline
$600$& $550$  & $42.3$ & $0.0431$
& $0.0259$& $0.0439$ \\ \hline
$840$  &  $550$ & $7900$ & $0.0001$
& $0.0001$& $0.1117$ \\ \hline
\end{tabular}
\end{center}
\end{table}
As we see from Table I, 
depending on the hierarchy of the dark matter masses,
the value of $\sigma_{0,123}$ has to be  tuned to obtain the 
observed  value of the total relic abundance.
We may say that the more hierarchical   the dark matter masses are,
the larger $\sigma_{0,123}$ is, and the larger $\Omega_{\chi_3} h^2$ is.
We then consider the asymmetric case, i.e.,
$\sigma_{0,123}\neq\sigma_{0,231}\neq\sigma_{0,312}$.
In Table II, we give some examples of this case
with fixed dark matter masses, 
$m_1=1000 $ GeV, $m_2=900 $ GeV and $m_3=550$ GeV,
where we have assumed that $m_1$ and $m_2$ are close, but
$m_3$ is about one half of $m_1$.
Since $\sigma_{0,123}$  is the size for 
the semi-annihilation $\chi_1\chi_2\to \chi_3 X$,
the relic abundance of $\chi_3$ is larger than the others  for larger $\sigma_{0,123}$.
Finally, we would like to point out that, 
since each semi-annihilation
produces a DM particle,
the semi-annihilation process can be few orders of magnitude larger than the standard process
where only the standard process exits (as we can see from Tables I and II).
The magnitude, of course, depends on a  model, but this can be
a useful information for model building.

\begin{table}
\caption{\footnotesize{The relic abundances for the  asymmetric case; i.e.,
$\sigma_{0,123}\neq\sigma_{0,231}\neq\sigma_{0,312}$,
 with $m_1=1000 $ GeV, $m_2=900 $ GeV and $m_3=550$ GeV.}
}
\begin{center}
\begin{tabular}{|c|c|c|c|c|c|} 
\hline
  $\sigma_{0,123}$ & $\sigma_{0,231}$ 
& $\sigma_{0,312}$ & $\Omega_{\chi_1} h^2$
& $\Omega_{\chi_2} h^2$& $\Omega_{\chi_3} h^2$\\ \hline
 $48.0$ & $2000.0$ 
& $48.4$ & $0.0325$
& $0.0007$& $0.0793$\\ \hline
$55.5$ & $65.0$ 
& $2000.0$ & $0.0003$
& $0.1118$& $0.0002$\\ \hline
$90.0$ & $1000.0$ 
& $100.3$ & $0.0121$
& $0.0011$& $0.0988$\\ \hline
 $110.0$ & $600.0$ 
& $145.2$ & $0.0067$
& $0.0015$& $0.1039$ \\ \hline
\end{tabular}
\end{center}
\end{table}

\section{A Model with three dark matter particles}
We extend the original radiative seesaw model of Ref.~\cite{Ma:2006km}
so as to have an additional discrete symmetry, $Z'_2$.
This can be done by introducing a SM singlet Majorana fermion
$\chi$ and a SM singlet real inert scalar $\phi$ in addition to the
inert Higgs doublet $\eta$ which is present in the original model.
The matter content of the model with the corresponding quantum numbers
is given in Table III.
\begin{table}
\caption{\footnotesize{The matter content of the model and
the corresponding quantum numbers. $Z_{2}
\times  Z'_2$ is  the unbroken discrete symmetry. 
The quarks are suppressed in the Table.}}
\begin{center}
\begin{tabular}{|c|c|c|c|c|} 
\hline
field & $SU(2)_L$ & $U(1)_Y$ & $Z_2$ &$Z'_2$	\\ \hline\hline
$(\nu_{Li},l_i)$	& $2$ 	& $-1/2$ 	& $+$ & $+$	\\ \hline
$l_{i}^c$ 		& $1$ 	& $1$		& $+$ & $+$	\\ \hline
$N^c_{i}$			& $1$		& $0$ 	& $-$ & $+$	\\ \hline
$H=(H^+,H^0)$& $2$ 	& $1/2$ 	& $+$ & $+$	\\ \hline
$\eta=(\eta^+,\eta^0)$& $2$ 	& $1/2$ 	& $-$ & $+$	\\ \hline
$ \chi $       & $1$    & $0$        & $+$ & $-$	\\ \hline
$ \phi $     & $1$    & $0$        & $-$ & $-$	\\ \hline
\end{tabular}
\end{center}
\end{table}

The $Z_2 \times Z'_2$ invariant Yukawa couplings of the lepton sector are given by
\begin{equation}
\mathcal{L}_Y 
= Y^e_{ij} H^\dag L_i  l_j^c
 + Y^{\nu}_{ik}L_i \epsilon \eta N_k^c
 + Y^\chi_k \chi N_k^c \phi
 + h.c. ~,
 \label{LY}
\end{equation}
and the Majorana mass terms of the right-handed neutrinos 
$N_k^c~(k=1,2,3)$ and the singlet fermion $\chi$ are\footnote{A similar model is considered in Ref.~\cite{SungCheon:2008ts}.}
\begin{equation}
\mathcal{L}_\mathrm{Maj} = \frac{1}{2} M_k N_k^c N_k^c + \frac{1}{2} M_\chi \chi^2 + h.c. 
\end{equation}
We may  assume without loss of generality that the right-handed neutrino mass matrix is diagonal and real.
As far as the light neutrino masses, which are generated radiatively, are concerned,
the last additional interaction  term in Eq.~(\ref{LY})  has no influence.
So the neutrino phenomenology is the same as in the original model.
The most general form of the $Z_2 \times Z'_2$-invariant  scalar potential 
can be written as
\begin{eqnarray}
V&=&m_1^2 H^\dag H + m_2^2 \eta^\dag \eta + \frac{1}{2} m_3^2 \phi^2
\nonumber \\&&
 + \frac{1}{2}\lambda_1 (H^\dag H)^2
 + \frac{1}{2}\lambda_2 (\eta^\dag \eta)^2
 + \lambda_3 (H^\dag H)(\eta^\dag \eta)
+\lambda_4 (H^\dag \eta)(\eta^\dag H)
 \nonumber \\&&
 + \frac{1}{2}\lambda_5 [(H^\dag \eta)^2+h.c.]
 + \frac{1}{4!}\lambda_6 \phi^4
 + \frac{1}{2}\lambda_7 (H^\dag H)\phi^2
 + \frac{1}{2}\lambda_8 (\eta^\dag \eta)\phi^2~,
 \label{potential}
\end{eqnarray}
from which we obtain the masses of the inert Higgs fields:
\begin{eqnarray}
m^2_{\eta^{\pm}}
&=&
m_2^2 + \lambda_3 v^2/2
\label{m1}\\
m^2_{\eta^0_R} &=&
m_2^2 + (\lambda_3 + \lambda_4 + \lambda_5)v^2/2
=m_2^2 + \lambda_Lv^2\label{m2}/2\\
m^2_{\eta^0_I} &=&
m_2^2 + (\lambda_3 + \lambda_4 - \lambda_5)v^2/2~,\label{m3}\\
m^2_\phi & =& m_3^2 + \lambda_7 v^2/2\label{m4}~.
\end{eqnarray}
Here, $\langle H \rangle =v/\sqrt{2}$ is the Higgs VEV,
and $\eta^0=(\eta^0_R+i\eta_I^0)/\sqrt{2}$. At this stage, we have assumed that
\be
\langle H\rangle &=&v/\sqrt{2}, \langle \eta\rangle = \langle \phi\rangle = 0~,
\label{VEV}
\ee
correspond to  the absolute minimum. (The sufficient condition for
the absolute minimum of Eq.~(\ref{potential}) is given below.)
As we can see from Table III, the cold DM candidates are
$N^c_k, \eta^0_R, \eta^0_I, \chi$ and $ \phi$,
where $\eta^0_R$ as dark matter
in the original model has been  discussed in detail in
Refs.~\cite{Barbieri:2006dq,LopezHonorez:2006gr,Dolle:2009fn}. 
To proceed, we assume that
the mass relations 
\be
M_k & >> & m_{\eta^\pm}, m_{\eta^0_I} >   m_{\eta^0_R} > m_{\phi},m_{\chi}
~\mbox{and}~m_{\eta^0_R} < m_{\phi}+m_{\chi}
\label{hr}
\ee
are satisfied.\footnote{The possibility $m_{\eta^0_I} <   m_{\eta^0_R}$
does not give any new feature of the model.}
These relations are chosen because  we would like  to meet the following
requirements:

\noindent
{\bf 1.}  \underline{$\mu\to e ~\gamma$}\\
The constraint coming from
$\mu\to e \gamma$  is given by \cite{Ma:2001mr}
\begin{eqnarray}
&&B(\mu\rightarrow e\gamma)={3\alpha\over 64
\pi(G_Fm_{\eta^\pm}^2)^2}
\left| ~\sum_k Y^\nu_{\mu k}Y^\nu_{ek}F_2
\left({M_k^2\over m_{\eta^\pm}^2}\right)
\right|^2 \lsim 2.4\times10^{-12}~,\label{mug}\\
&&F_2(x)={1\over 6(1-x)^4}(1-6x+3x^2+2x^3-6x^2\ln x)~,\nonumber
\end{eqnarray}  
where the upper bound is taken from
Ref.~\cite{Hayasaka:2010et}.
A similar, but slightly weaker bound for $\tau \to \mu (e) \gamma$ given
in Ref.~\cite{Hayasaka:2010et}
has to be satisfied, too.
Since $F_2(x) \sim 1/3x$ for $x >>1$, while  
$1/12 < F_2(x) < 1/6 $ for $0< x <1$, the constraint 
can be readily  satisfied if
$M_{k}<< m_{\eta^\pm}$ or $M_{k}>> m_{\eta^\pm}$.\\

\noindent
{\bf 2.}  \underline{$g_\mu-2$}\\
The extra contribution  to the anomalous magnetic
moment of the muon, $a_\mu=(g_\mu-2)/2$, is given by \cite{Ma:2001mr} 
\begin{eqnarray}
\delta a_\mu &=& {\frac{m_\mu^2}{16 \pi^2 m_{\eta\pm}^2}}
\sum_k Y^\nu_{\mu k}Y^\nu_{\mu k}F_2
\left({M_k^2\over m_{\eta^\pm}^2}\right)~ .
\end{eqnarray}  
If we assume that 
$|\sum_k Y^\nu_{\mu k}Y^\nu_{\mu k} F_2
\left({M_k^2\over m_{\eta^\pm}^2}\right)|\simeq
|\sum_k Y^\nu_{\mu k}Y^\nu_{ek} F_2
\left({M_k^2\over m_{\eta^\pm}^2}\right)|$,
then we  obtain 
\be
|\delta a_\mu | &\simeq & 2.2\times 10^{-5} 
B(\mu\rightarrow e\gamma) \lsim 3.4 \times 10^{-11}
\ee
if Eq.~(\ref{mug}) is satisfied,
where the upper bound is taken from
Ref.~\cite{Nakamura:2010zzi}. So, the constraint from $a_\mu$ has no significant 
influence.\\

\noindent
{\bf 3.} \underline{Stable and global minimum}\\
The DM stabilizing symmetry $Z_2$ remains unbroken if
\be
& & m_1^2 < 0~,~m_2^2 > 0~,~m_3^2 > 0~,\nn\\
&  &\lambda_1~,~\lambda_2~,~\lambda_6 > 0~,~
\lambda_3+\lambda_4-|\lambda_5|,\lambda_3 >
 -\frac{1}{2}(\lambda_1\lambda_2)^{1/2}~,\nn\\
& &\lambda_7> -\frac{1}{2}(\lambda_1\lambda_6/3)^{1/2}~,~
\lambda_8> -\frac{1}{2}(\lambda_2\lambda_6/3)^{1/2}~,
\label{cnd3}
\ee
are satisfied. These conditions are sufficient for Eq.~(\ref{VEV})
to correspond to the absolute minimum.
Since $m_{\eta^0_R}^2-m_{\eta^0_I}^2=
\lambda_5 v^2$, a negative $\lambda_5$  is consistent with Eq.~(\ref{hr}).\\

\noindent
{\bf 4.} \underline{Electroweak precision}\\
The electroweak precision measurement requires
\cite{Barbieri:2006dq,Nakamura:2010zzi}
\be
\Delta T &\simeq & 0.54
\left(\frac{m_{\eta^\pm}-m_{\eta^0_R}}{v}\right)
\left(\frac{m_{\eta^\pm}-m_{\eta^0_I}}{v}\right)
=0.02_{-0.12}^{+0.11}~, 
\label{cnd4}
\ee
for $m_h=115-127$ GeV.
Therefore,  
$|m_{\eta^\pm}-m_{\eta^0_R}|~,~
|m_{\eta^\pm}-m_{\eta^0_I}| \lsim 100~\mbox{GeV}$ 
is sufficient to meet the requirement.

Then, with the assumption of the above mass relations, we look at the
radiative neutrino mass matrix \cite{Ma:2006km}:
\be
({\cal M}_\nu)_{ij} 
&=&\sum_k \frac{Y^\nu_{ik} Y^\nu_{jk} M_k}{16 \pi^2}
\left[~
\frac{m_{\eta_R^0}^2}{m_{\eta_R^0}^2-M_k^2}
\ln\left(\frac{m_{\eta_R^0}}{M_k}\right)^2
-\frac{m_{\eta_I^0}^2}{m_{\eta_I^0}^2-M_k^2}
\ln\left(\frac{m_{\eta_I^0}}{M_k}\right)^2
~\right]~\label{Mnu}\\
&\simeq &-\sum_k \frac{Y^\nu_{ik} Y^\nu_{jk} }{16 \pi^2}
\left[
\frac{m_{\eta_R^0}^2}{M_k}\ln\left(\frac{m_{\eta_R^0}}{M_k}\right)^2
-\frac{m_{\eta_I^0}^2}{M_k}\ln\left(\frac{m_{\eta_I^0}}{M_k}\right)^2
\right]~\mbox{for}~m_{\eta_R^0},m_{\eta_I^0} << M_k~.\nn
\ee
Since $({\cal M}_\nu)_{ij}$ are of order $ 10^{-1}$ eV
and $m_{\eta_R^0}^2-m_{\eta_I^0}^2=\lambda_5 v^2$,
we need $\sum_k Y^\nu_{ik} Y^\nu_{jk} \lsim O(10^{-9})$ for 
 $|\lambda_5|\gsim O(0.1)$.
Note, however, that 
this does not automatically imply that $ \sum_{i,k}^3 |Y^\nu_{ik}|^2
\lsim O(10^{-9})$; and in fact, it could be much larger 
 if we assume a specific flavor structure of $Y^\nu_{jk}$.
 If there exists another source for the neutrino mass matrix, we have to add it
to Eq.~(\ref{Mnu}).

\subsection{Relic abundance of dark matter}
Now we come to the relic abundance of DM.
Under  the assumption about the mass relations [Eq.~(\ref{hr})], we have to consider
 the following 
annihilation processes:\footnote{We neglect the coannihilations, such as that of $\eta_R^0$ with
$\eta_I^0$ and $\eta^\pm$ below.}
\be
& &\bullet \eta^0_R ~\eta^0_R \leftrightarrow \mbox{SMs}~,~
\bullet \phi ~\phi  \leftrightarrow \mbox{SMs}~
(\mbox{Standard annihilation})\\
& &
\bullet\eta^0_R~ \eta^0_R \leftrightarrow \phi ~\phi ~,~
\bullet \chi ~\chi\leftrightarrow  \phi~ \phi ~
(\mbox{Conversion})\\
& &
\bullet\eta^0_R ~\chi \leftrightarrow \phi ~\nu_L~,~
\bullet\chi ~\phi \leftrightarrow \eta^0_R ~\nu_L~,~
\bullet\phi ~\eta^0_R \leftrightarrow \chi ~\nu_L~
(\mbox{Semi-annihilation})
\ee

\begin{figure}[tb]
\begin{center}
  \includegraphics[clip]{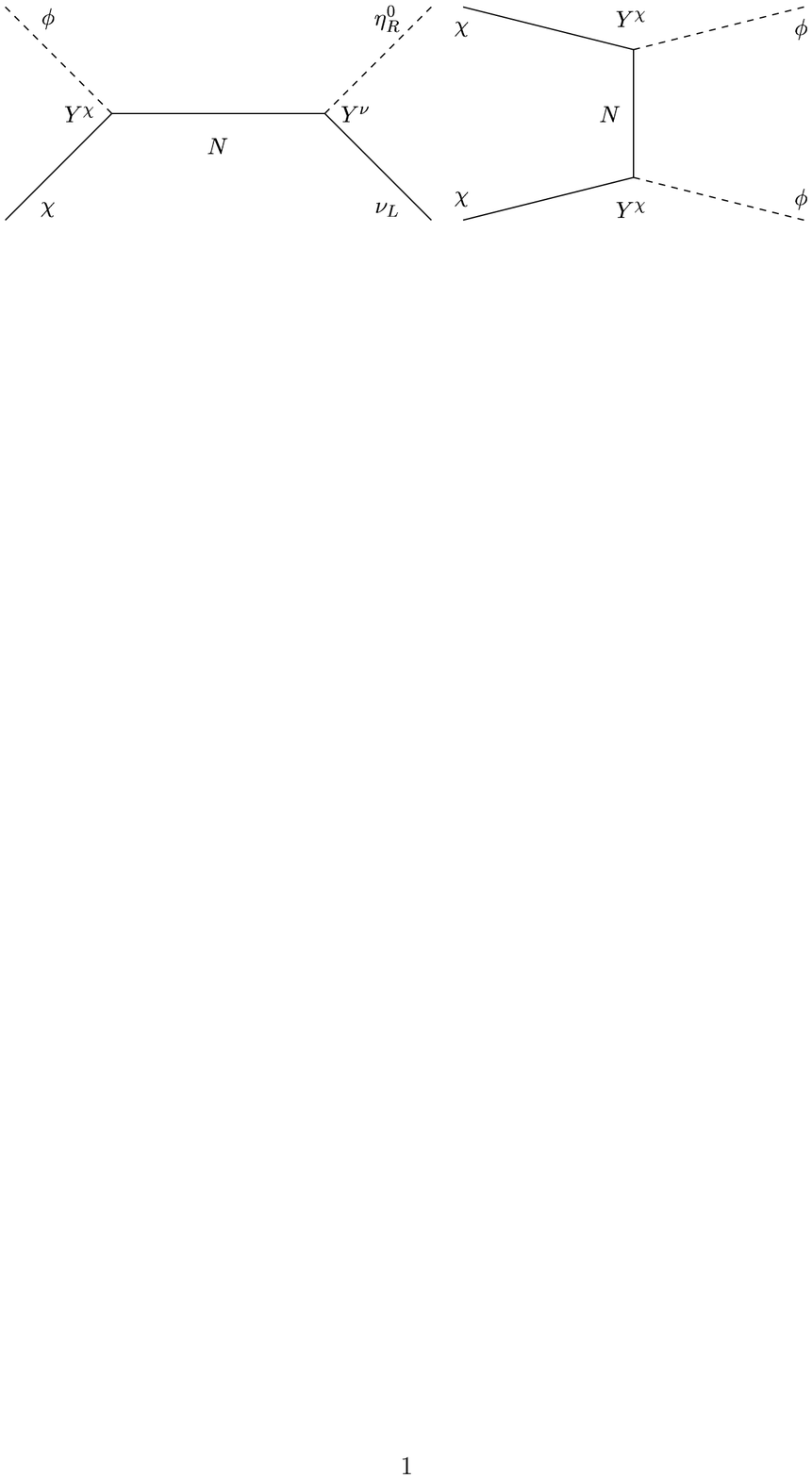}
\end{center}
\caption[]{\label{semi1} Semi-annihilation diagram (left)
and conversion (right).
}
\end{figure}
\begin{figure}
  \includegraphics[width=8cm]{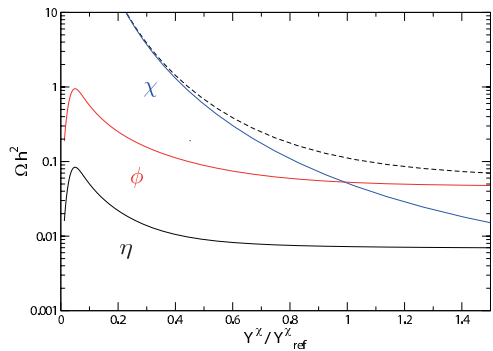}
\caption{\label{semi-Y}\footnotesize
$Y^\chi$ dependence of the relic abundances,
$\Omega_T h^2$ (dashed curve), 
$\Omega_\eta h^2$ (black curve), $\Omega_\chi h^2$ (blue (light gray) curve), 
$\Omega_\phi h^2$ (red (gray) curve), 
where $Y^\chi$ controls the size of the 
semi-annihilation and conversion shown in Fig.~\ref{semi1}.
The input parameter values are given in Eq.~(\ref{input3}).
}
\end{figure}

\noindent
We have yet not specified the relative size of $m_\chi$ and $m_\phi$.
If $\chi$ is lighter than $\phi$, the  conversion  of $\chi$ into $\phi$ is 
kinematically forbidden, and the semi-annihilation in Fig.~\ref{semi1} is the only
kinematically allowed annihilation for $\chi$.
So, we will consider below only the case $m_\chi > m_\phi$.
First,  we  consider
a benchmark set 
of the input parameter values:
\be 
m_{\eta_R^0} &= &200 ~\mbox{GeV}~,~
m_\chi = 190 ~\mbox{GeV}~,~
m_\phi = 180 ~\mbox{GeV}~,\nn\\~
m_{\eta^\pm} &= &m_{\eta_I^0}=210 ~\mbox{GeV}~,\nn\\
m_h &=& 125~\mbox{GeV}~,~M_1 = M_2=M_3= 1000 ~\mbox{GeV}~,
\label{input3}\\
\lambda_3 &=&  -0.065~,~\lambda_7= 0.1~,~\lambda_8= 0.1~,~
\lambda_L=-0.2~,\nn\\
\sum_{k=1}^3 |Y^\chi_k|^2&=&3(0.7)^2~,~
\sum_{i,k=1}^3 |Y^\nu_{ik}|^2=9(0.01)^2~.\nn
\ee
 With this choice of the parameter values, we obtain
 \be
<\sigma(\eta_R^0 \eta_R^0; \mbox{SMs})v >&=&
45.66-38.21/x~,~
<\sigma(\phi \phi; \mbox{SMs})v >=
5.93-1.92/x~~,\nn\\
<\sigma(\eta_R^0 \eta_R^0; \phi \phi)v >&=&
0.46+0.29/x~,~
<\sigma(\chi \chi; \phi\phi)v >=
0+77.18/x~,\\
<\sigma(\chi \eta_R^0; \phi\nu_L)v >&=&
0.02+0.01/x~,~
<\sigma(\eta_R^0 \phi; \chi\nu_L)v >=
0.07+0.02/x~,\nn\\
<\sigma(\chi \phi; \eta_R^0\nu_L)v >&=&
0.07+0.04/x,\nn
\ee
in units of $10^{-9}~\mbox{GeV}^{-2}$, and
\be
 \Omega_Th^2 &=&0.1094~,~\Omega_\eta h^2 =0.0062~,~\Omega_\chi h^2 =
0.0511~,~\Omega_\phi h^2 = 0.0521~,
\ee
where $x=(1/m_{\eta_R^0}+1/m_{\chi}+1/m_{\phi})^{-1}/T=\mu/T$.
As we see from Fig.~\ref{semi1}, the size of the 
semi-annihilation and conversion is controlled  by $Y^\chi_k$.
In Fig.~\ref{semi-Y}, we show the $Y^\chi$ dependence
of the individual abundances, where we have varied $\sum_k|Y^\chi_k|^2$, and
$Y^\chi/Y^\chi_{\rm ref}$ stands 
for 
$\left(\sum_k|Y^\chi_k|^2 /3 (0.7)^2\right)^{1/2}$.
If $Y^\chi/Y^\chi_{\rm ref} <<1$, the conversion of $\chi$ and the semi-annihilations $\chi ~\phi \to \eta_R^0~\nu_L~,~
\chi ~\eta_R^0 \to \phi~\nu_L$ become small, such that $\Omega_\chi h^2$ in particular  increases.

\begin{figure}[htb]
\begin{center}
  \includegraphics[clip]{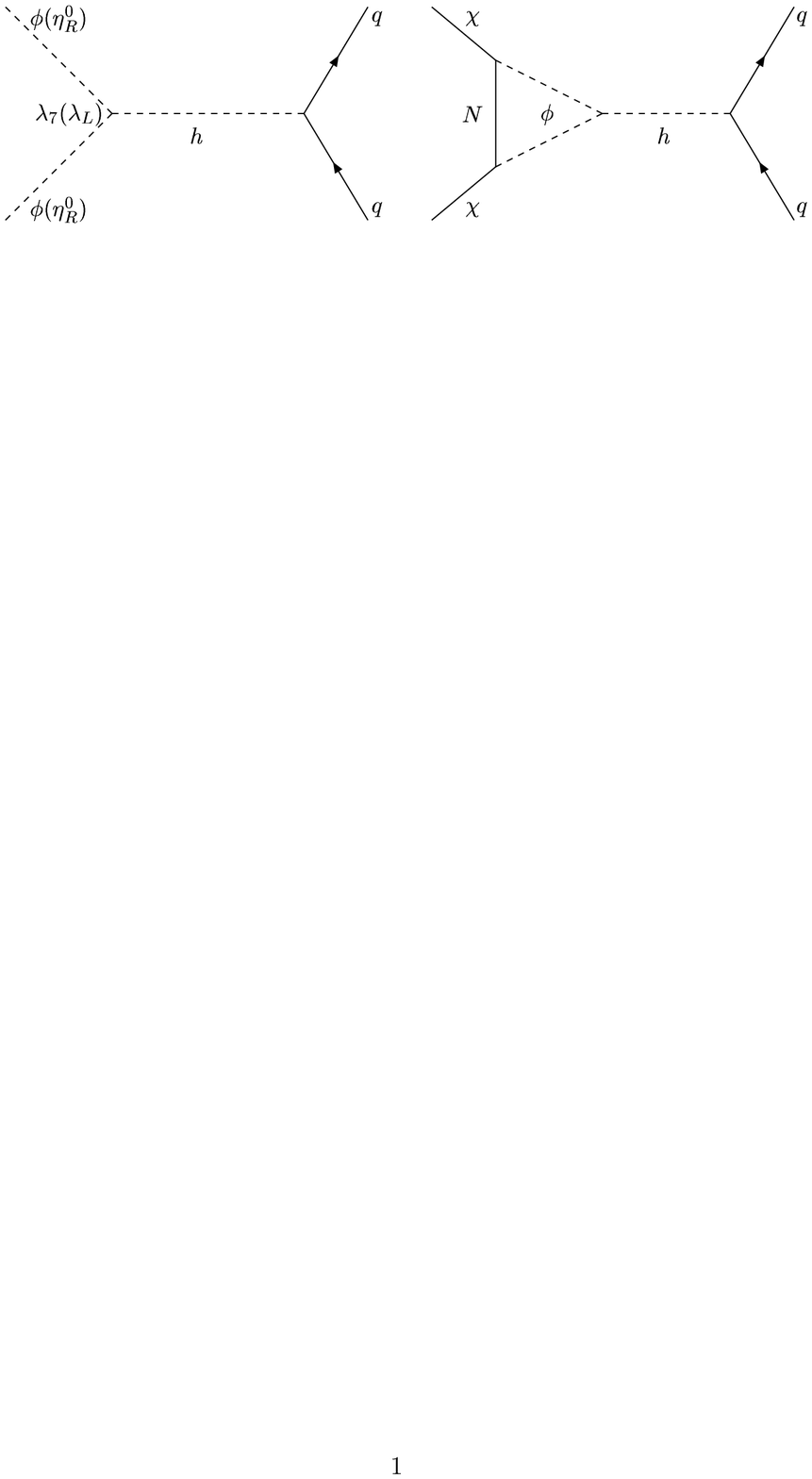}
\end{center}
\caption[]{\label{diagram-q}Tree (left) and one-loop (right) level interactions with the quarks.}
\end{figure}

\subsection{Imposing constraints}

To be more realistic, we have to impose constraints from 
direct detection, collider experiments,
and perturbativity, $|\lambda_i|, |Y_{ij}^\nu |, |Y^\chi_i| < 1$,
in addition to Eqs.~(\ref{mug})--(\ref{cnd4}),
which we shall do next.
The DM particles $\phi$ and $\eta_R^0$ have tree-level interactions to the quarks,
which are shown in  Fig.~\ref{diagram-q}.\footnote{
Direct detection of two DM particles has been discussed,
for instance, in Refs.~\cite{Hur:2007ur,Feldman:2010wy,Batell:2010bp,Belanger:2011ww}.
LHC signals of $\eta$ dark matter have been discussed
in Refs.~\cite{Barbieri:2006dq,Dolle:2009ft,Gustafsson:2012aj}. 
See also Refs.~\cite{SungCheon:2008ts,Batell:2010bp}.}
In the following discussions, we ignore the one-loop contributions 
such as the right diagram in Fig.~\ref{diagram-q}.\footnote{
There exist also one-loop corrections to $\eta_R^0 q
\to \eta_R^0 q$ \cite{Cirelli:2005uq}. See also Ref.~\cite{Schmidt:2012yg}.}
The spin-independent elastic cross section off the nucleon 
$\sigma(\phi(\eta^0_R))$
is given by \cite{Barbieri:2006dq}
\be
\sigma(\phi(\eta^0_R))
&=&\frac{1}{4\pi} 
\left( \frac{\lambda_{7(L)} \hat{f}
m_N }{m_{\phi(\eta_R^0)} m_h^2} \right)^2
\left(\frac{m_N m_{\phi(\eta_R^0)}}{m_N+m_{\phi(\eta_R^0)}}
\right)^2~,
\ee
where $m_N$ is the nucleon mass, and
$\hat{f}\sim 0.3$ stems from the nucleonic matrix element \cite{Ellis:2000ds}.
The cross sections  have to satisfy
\be
\left(\frac{\sigma(\phi)}{\sigma_{\rm UB}(m_\phi)}\right)
\left(\frac{\Omega_\phi h^2}{0.112}\right)+
\left(\frac{\sigma(\eta_R^0)}{\sigma_{\rm UB}(m_{\eta_R^0})}\right)
\left(\frac{\Omega_\eta h^2}{0.112}\right) \lsim 1~,
\ee
where $\sigma_{\rm UB}(m)$ is the current experimental limit
on the cross section for the DM mass $m$.

In the absence of $\chi$ and $\phi$, the lower-mass region
$60 ~\mbox{GeV} \lsim m_{\eta_R^0} \lsim 80~\mbox{GeV}$
is consistent  with all the experimental constraints  
\cite{Dolle:2009fn,Gustafsson:2012aj}.\footnote{There exists a marginal possibility to expend 
slightly this upper bound \cite{LopezHonorez:2010tb}.}
But the elastic cross section
$\sigma(\eta_R^0) \simeq 7.9 \times 10^{-45}
(\lambda_L/0.05)^2 (60~\mbox{GeV}/m_{\eta_R^0})^2~\mbox{cm}^2$
with  $\lambda_L \gsim 0.05$ in this mass range  
may exceed the upper bound of 
the XENON100 result \cite{Aprile:2011hi},\footnote{See also \cite{Angloher:2008jj}--\cite{Kim:2012rz}.} 
$ 7.0\times 10^{-45}~\mbox{cm}^2$ for the DM mass $50$ GeV
at $90 \%$ C.L.
The higher-mass region, i.e.,
$ m_{\eta_R^0}\gsim 500 ~\mbox{GeV} $ with $\sigma(\eta^0_R )
\simeq 4.6 \times 10^{-46}
(\lambda_L/0.1)^2 (500~\mbox{GeV}/m_{\eta_R^0})^2~\mbox{cm}^2$, 
will  be significant for next-generation experiments 
such as 
SuperCDMS \cite{Bruch:2010eq}, XENON1T \cite{Selvi:2011zz} or
XMASS \cite{Sekiya:2010bf}.
\begin{figure}
  \includegraphics[width=8cm]{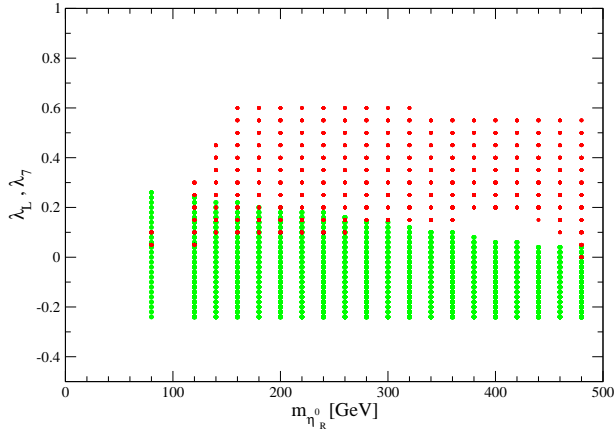}
\caption{\label{lambda-m}\footnotesize
The allowed regime in the $\lambda_L
(\lambda_7)-m_{\eta_R^0}$  plane
for $(\delta_1=10~,~
\delta_2=10)$ GeV with
$m_\chi=m_{\eta_R^0}-10~\mbox{GeV}$,
$m_\phi=m_{\eta_R^0}-20~\mbox{GeV}$ and 
$M_k=1000$ GeV. The green(light gray) and red (dark gray) points are for $\lambda_L$ and
$\lambda_7$, respectively.
}
\end{figure}

The presence of $\chi$ and $\phi$ changes the situation.
Firstly, the separation of two   allowed regions of $m_{\eta_R^0}$ 
disappears:  As far as the relic abundance
is concerned, $m_{\eta_R^0}$ is allowed  between $80$ and $500$ GeV too,
as we have seen, because  $\chi$ and $\phi$  also contribute to
the relic abundance.
Secondly, the parameter space becomes considerably larger.
To see how the allowed parameter space of the model without $\chi$ and $\phi$
changes, we consider  a set of $(\delta_1=m_{\eta^\pm}-m_{\eta_R^0}~,~
\delta_2=m_{\eta_I^0}-m_{\eta_R^0})$, for which the
allowed parameter space without $\chi$ and $\phi$  is very small.
For  $(\delta_1=10~,~
\delta_2=10)$ GeV, for instance, there is no allowed 
range of $m_{\eta_R^0} \lsim 500$ GeV \cite{Dolle:2009fn}; 
the low-mass  range of $m_{\eta_R^0}$, for  which
the relic abundance $\Omega_\eta h^2$ is consistent,
does not satisfy the LEP constraint. 
Below we show how this situation changes in the presence of 
 $\chi$ and $\phi$. The LEP constraint implies
that the region satisfying $m_{\eta_R^0} \lsim 80$ GeV
and $m_{\eta_I^0} \lsim 100$ GeV
with $\delta_2 \gsim 8$ GeV is excluded \cite{Dolle:2009fn}.
Therefore, for  $(\delta_1=10~,~
\delta_2=10)$ GeV, we have to consider 
only $m_{\eta_R^0} > 80$ GeV.
Further, to suppress the parameter space,
we assume that 
$m_\chi=m_{\eta_R^0}-10~\mbox{GeV}$,
$m_\phi=m_{\eta_R^0}-20~\mbox{GeV}$, and 
$M_k=1000$ GeV, and we scan $m_{\eta_R^0}$ 
from $80$ to $500$ GeV.

Figure~\ref{lambda-m} shows the allowed area in the $\lambda_L
(\lambda_7)-m_{\eta_R^0}$  plane, where
 all the constraints are taken into account.
The allowed mass range
for 
$m_{\eta_R^0}$ is extended as expected.
The reason that
there are no allowed points around $m_{\eta_R^0}\simeq 100$ GeV
is the following:
Since we keep  the mass difference fixed,
we have $m_\phi=m_{\eta_R^0}-20\simeq 80$ GeV there.
This is  the threshold regime
for the process $\phi \phi\to W^+ W^-$.
So, for $m_{\eta_R^0}$ just below $100$ GeV,
the annihilation cross section for $\phi$ is 
small because of small $\lambda_7$ in this range of
$m_\phi$, and therefore the relic abundance $\Omega_\phi h^2$ exceeds $0.12$.
We see that  $m_{\eta_R^0}=80$ GeV is allowed, on the other hand.
This allowed area exists, though $\lambda_7$ is small,
because around $m_{\phi}=62$ GeV, the s-channel annihilation of $\phi$ becomes
resonant due to $m_h=125$ GeV.
For $m_{\eta_R^0}$ just above $100$ GeV,
the annihilation cross section for $\phi$ is 
large because the channel to $W^+ W^-$ is now open, so that $\Omega_\phi h^2$ cannot supplement the anyhow small $\Omega_\eta h^2$.

\begin{figure}
  \includegraphics[width=8cm]{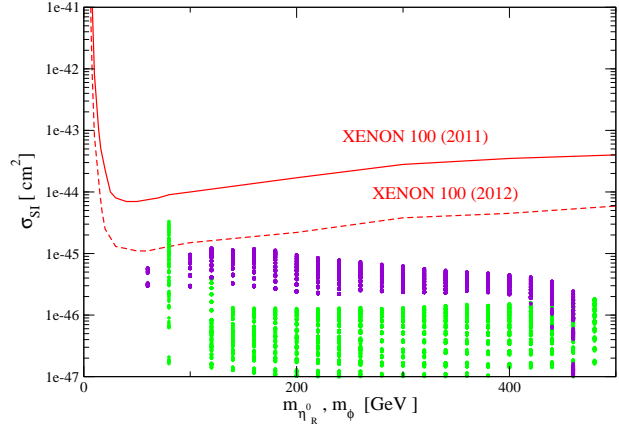}
\caption{\label{direct}\footnotesize
The spin-independent cross section
off the nucleon  is plotted as a function of the DM mass.
The green (light gray) and violet (dark gray) areas  are for $\eta$ and
$\phi$ DM's, respectively, where we have used
$(\delta_1=10~,~
\delta_2=10)$ GeV with
$m_\chi=m_{\eta_R^0}-10~\mbox{GeV}$,
$m_\phi=m_{\eta_R^0}-20~\mbox{GeV}$ and 
$M_k=1000$ GeV. 
}
\end{figure}

 If we suppress
the constraint from the direct detection, we have a prediction
on the direct detection.
Figure~\ref{direct} shows the spin-independent cross section off the nucleon
versus the DM mass;
the green (light gray) does so for the $\eta$ DM, and the violet (dark gray) area for the $\phi$ DM.
We see that the the spin-independent cross sections not only are consistent
with the current bound of XENON100 \cite{Aprile:2011hi}, but also are within the accessible range of future experiments.

\subsection{Indirect detection}
If DM annihilates sufficiently into SM particles,
the resulting cosmic rays may be observable.
These are indirect signals of DM, and 
in fact excesses in  $e^+$
\cite{Adriani:2008zr}--\cite{FermiLAT:2011ab} and 
in  $\gamma$ \cite{Aharonian:2004wa}--\cite{Acciari:2010ab}
have been  recently reported.
Indirect detection
of DM has been studied within the framework of a two-component DM system in 
 Refs.~\cite{Boehm:2003ha,Cao:2007fy,Huh:2008vj,Fairbairn:2008fb,Zurek:2008qg,Feldman:2010wy,Fukuoka:2009cu,Fukuoka:2010kx}, 
  and also within the inert Higgs model
 in Refs.~\cite{LopezHonorez:2006gr,Gustafsson:2007pc,Agrawal:2008xz,Suematsu:2010gv}.
As we see from the semi-annihilation diagram in Fig.~\ref{semi1}, the process
produces only a left-handed neutrino as the SM particle.
Therefore,  we are particularly interested in
the neutrinos from the annihilation of captured DM  
in the Sun \cite{Silk:1985ax}--\cite{Hooper:2002gs}
(see Refs.~\cite{Jungman:1995df,Bertone:2004pz} for a review,
and Refs.~\cite{ Agrawal:2008xz,Andreas:2009hj} for the case of the inert Higgs model),
because (i) the semi-annihilation  produces a monochromatic neutrino
($E_\nu\simeq m_{\eta^0_R}+m_\phi-m_\chi$, for instance)
in addition to those with $E_\nu\simeq m_{\eta^0_R}$
along with the  continuum spectrum, 
(ii) these neutrinos can be observed at neutrino telescopes 
\cite{Tanaka:2011uf,IceCube:2011aj,Zornoza:2012tu},
and (iii) the evolution equations 
of the DM numbers in the Sun,
which describe approaching
 equilibrium between the capture  and annihilation 
 (including conversion and semi-annihilation) rates of DM,
are coupled now.

We denote the number of DM particles $ \eta, \chi$ and $\phi$
in the Sun by $N_i$, with $i=\eta, \chi$ and $ \phi$, respectively.
Then the change of $N_i$ with respect to time $t$ is given by
\be
\dot{N}_\eta &=&
C_\eta-C_A(\eta \eta \leftrightarrow \mbox{SM}) N_\eta^2
-C_A(\eta \eta \leftrightarrow \phi \phi)  N_\eta^2
-C_A(\eta \chi\leftrightarrow \phi \nu_L) N_\eta  N_\chi \nn\\
& &
-C_A(\eta \phi\leftrightarrow \chi \nu_L)N_\eta N_\phi
+C_A(\phi \chi\leftrightarrow \eta\nu_L) N_\chi N_\phi~,\\
\dot{N}_\chi&=&
C_\chi
-C_A(\chi \chi \leftrightarrow \phi \phi)  N_\chi^2
-C_A(\eta \chi\leftrightarrow \phi \nu_L) N_\eta  N_\chi \nn\\
& &
+C_A(\eta \phi\leftrightarrow \chi \nu_L)N_\eta N_\phi
-C_A(\phi \chi\leftrightarrow \eta\nu_L) N_\chi N_\phi~,\\
\dot{N}_\phi &=&
C_\phi-C_A(\phi \phi \leftrightarrow \mbox{SM}) N_\phi^2
+C_A(\eta \eta \leftrightarrow \phi \phi)  N_\eta^2
+C_A(\chi \chi \leftrightarrow \phi \phi)  N_\chi^2\nn\\
& &+C_A(\eta \chi\leftrightarrow \phi \nu_L) N_\eta  N_\chi 
-C_A(\eta \phi\leftrightarrow \chi \nu_L)N_\eta N_\phi
-C_A(\phi \chi\leftrightarrow \eta\nu_L) N_\chi N_\phi~,
\ee
where the $C_i$'s are  the capture rates in the Sun, and the $C_A$'s are proportional to the annihilation
cross sections times the relative DM velocity per volume in the
limit $v\to 0$:
\be
C_{\phi(\eta)}&\simeq & 1.4 \times 10^{20}
f(m_{\phi(\eta_R^0)} )\left(\frac{\hat{f}}{0.3}\right)^2
 \left(\frac{\lambda_{7(L)}}{0.1}\right)^2 
\left[\frac{m_h}{125~\mbox{GeV}}\right]^{-4}\nn\\
& \times& 
\left( \frac{200~\mbox{GeV}}{m_{\phi(\eta_R^0)}} \right)^2
\left(\frac{\Omega_{\phi(\eta)}h^2}{0.112}\right) \mbox{s}^{-1}~,
~C_\chi=0~,
\ee
where the function $f(m_{\phi(\eta_R^0)} )$ 
depends on the form factor of the nucleus,
elemental abundance, kinematic suppression 
of the capture rate, etc., varying from $O(1)$ to $O(0.01)$
depending  on the DM mass
\cite{Kamionkowski:1991nj,Kamionkowski:1994dp}. 
The annihilation rates, $C_A$, can be calculated
from \cite{Griest:1986yu}
\be
C_A(ij \leftrightarrow \bullet) &=& 
\frac{<\sigma(i j ;\bullet) v>}{V_{ij}}~,~
V_{ij}=5.7\times 10^{27} \left( \frac{100~\mbox{GeV}}{\mu_{ij}}
 \right)^{3/2} \mbox{cm}^3\, ,
\ee
with
$ \mu_{ij}=2 m_i m_j /(m_i+m_j)$ in the limit $v\to 0$.

\begin{figure}
  \includegraphics[width=8cm]{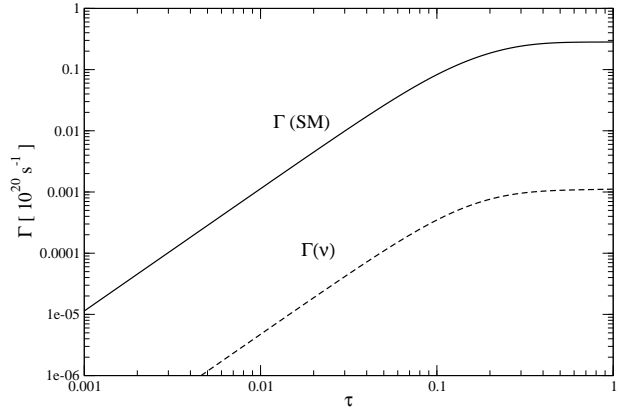}
\caption{\label{nu-rate}\footnotesize
The time evolution of the annihilation rates
$\Gamma(\mbox{SM})$ 
and $\Gamma(\nu)$,
where $\tau=t/t_\odot$, and
the input parameter values are given in  Eq. (\ref{input3}).}
\end{figure}

There are fixed points of  the evolution equations
which correspond to  equilibrium.
Since at the time of the sun's birth the numbers $N_i$  were zero,
the $N_i$'s increase with time and 
approach the fixed-point values, i.e., equilibrium, at which point $N_i$  assumes its maximal
value.
So, the question is whether the age of the Sun, $t_\odot \simeq 4.5\times 10^9$ years,
is old enough for $N_i$ to reach  equilibrium.
We see from the evolution equations that
the fixed-point values are roughly proportional to $(C_i /C_A)^{1/2}$,
implying that we need large capture rates $C_i$ to obtain large $N_i(t_\odot)$.
The annihilation, conversion and semi-annihilation rates at $t=t_\odot$ are given by
\be
\Gamma(ij;\bullet ) &=& d_{ij}C_A(ij \leftrightarrow \bullet) 
N_i (t_\odot) N_j(t_\odot) ~,
\ee
where $d_{ii}=1/2$ and $d_{i j}=1$ for $i\neq j$.
The observation rate of the neutrinos, $\Gamma_{\rm detect}$, is proportional to
$\Gamma(ij;\bullet )$. 
As a benchmark, we use the same set 
of the input parameter values as in Eq.~(\ref{input3}).
 In Fig.~\ref{nu-rate}, we show the time evolution of\footnote{For the monochromatic neutrinos, i.e. $ \Gamma(\nu) $, we 
 have added all the semi-annihilations, because for the mass values 
 given in Eq.~(\ref{input3}) the neutrino energies are
 all close to $200$ GeV. Moreover, 
 the first term in the r.h.s. of Eq.~(\ref{gamma-nu}) (which counts  neutrinos of
 $m_{\eta_R^0}+m_\phi-m_\chi=
 190$ GeV)
 is a dominant contribution with about $95$\%.
  }
\be
\Gamma(\mbox{SM}) &=&
C_A(\eta \eta \leftrightarrow \mbox{SM}) N_\eta^2/2
+C_A(\phi \phi \leftrightarrow \mbox{SM}) N_\phi^2/2~,\\
\Gamma(\nu) &=&
C_A(\eta \phi\leftrightarrow\chi\nu) N_\eta N_\phi
+C_A(\eta \chi\leftrightarrow \phi\nu) N_\eta N_\chi
+C_A(\chi \phi\leftrightarrow \eta\nu) N_\chi N_\phi~,
\label{gamma-nu}\\
\Gamma(\nu\nu) &=&
C_A(\eta \eta \leftrightarrow \nu\nu) N_\eta^2/2~,
\ee
scaled to $10^{20}~\mbox{s}^{-1}$,  as function of  $\tau=t/t_\odot$.
As we see from Fig.~\ref{nu-rate}, the  rates seem to be saturated:
$\Gamma(\mbox{SM})$ is in fact saturated,
but $\Gamma(\nu)$ does not reach its fixed-point value of
$0.002 \times 10^{20}~\mbox{s}^{-1}$ at $\tau=t/t_\odot=1$.
The saturated value of $\Gamma(\mbox{SM})$ is 
$0.28\times 10^{20}~\mbox{s}^{-1}$ for the 
 input parameters of Eq.~(\ref{input3}),
 which is consistent with the upper limit of
$\sim 2.73 \times 10^{21}~\mbox{s}^{-1}$ 
for $m_{\rm DM}=250~\mbox{GeV}$
of  
the AMANDA-II / IceCube neutrino
telescope  \cite{IceCube:2011aj}.
As for the monochromatic neutrinos, we obtain
$\Gamma(\nu)=1.1\times 10^{-3}\times 10^{20}~\mbox{s}^{-1}$
and $\Gamma(\nu\nu)=
1.3 \times 10^{-7}\times 10^{20}~\mbox{s}^{-1}$. 
To estimate the detection rate   $\Gamma_{\rm detect}$, we use
the formula 
 \cite{Halzen:2010yj}
 \be
  \Gamma_{\rm detect} = A P(E_\nu) \Gamma_{\rm inc}~ ,
  \ee
  where $A$ is the detector area facing 
  the incident beam, $P(E_\nu)$ 
  is the probability for detection as a function of the neutrino energy $E_\nu$,
  and  $\Gamma_{\rm inc}=\Gamma/4 \pi R_\odot^2$ is 
the incoming neutrino flux
--- i.e., the number of neutrinos per unit area per second on the Earth
 (where $R_\odot$ is the distance to the Sun $\simeq  1.5\times 10^8$ km).\footnote{A  sophisticated method to compute the observation rates
  at IceCube was recently developed in \cite{Scott:2012mq}.}
 The probability $P(E_\nu)$ may be approximated as 
the ratio of the  effective detector  length $L$ to the mean free path 
of the neutrinos in the detector.
For the neutrinos (anti-neutrinos), one finds
$P(E_{\nu(\bar{\nu})})\simeq 2.0 (1.0)
\times 10^{-11}(L/\mbox{km})(E_{\nu(\bar{\nu})}/\mbox{GeV})$,
arriving at
\be
  \Gamma_{\rm detect}
&  \simeq &
2.2 (1.1)\times 10^{-21}\left(\frac{A}{\mbox{km}^2}\right)
\left(\frac{L}{\mbox{km}}\right)
\left(\frac{E_{\nu(\bar{\nu})}}{\mbox{GeV}}\right)~
\left(\frac{\Gamma}{\mbox{s}^{-1}}\right)~
\mbox{yr}^{-1}~,
\ee
which implies  that, for  the input parameters
of Eq.~(\ref{input3}), $0.05$ events of 
monochromatic neutrinos with $\sim 200$ GeV per year 
may be detected at IceCube \cite{IceCube:2011aj}, where
we have used $A=1\mbox{km}^2$, $L=1\mbox{km}$.

 A total of $0.05$ events per year 
 may be too small to be  realistic.
 However, we would like to note that we have studied only one point in the whole parameter space.
 It will be our future program to implement the 
  sophisticated method of Ref.~\cite{Scott:2012mq} and to 
survey  the whole parameter space. How to observe the monochromatic neutrinos 
 at neutrino telescopes should also be addressed \cite{Esmaili:2009ks}.
Finally, we would like to note that if at least one of the fermionic DM particles in a multi-component
DM system has odd parity of the discrete lepton number,
then a monochromatic left-handed neutrino,
which is also odd,
 can be produced together with this fermionic  DM in a semi-annihilation of DM particles.

\section{Conclusion}

We have considered the conversion and  semi-annihilation
of DM  in a multi-component  DM system. 
 We have found in  fictive models that
 these non-standard DM annihilation processes
 can influence
 the  relic abundance of DM a lot,
which has been  recently  found for 
 two-component DM systems
in Refs.~\cite{D'Eramo:2010ep,Belanger:2011ww,Belanger:2012vp}. 

As a concrete three-component DM system, we have considered
a radiative seesaw model of Ref.~\cite{Ma:2006km},  which is extended
to include an extra Majorana fermion $\chi$ and an extra real scalar boson $\phi$.
 The DM stabilizing symmetry is promoted to $Z_2\times Z'_2$,
 and we have assumed that 
$\eta_R^0$ (the CP-even neutral component 
of the  inert Higgs $SU(2)_L$ doublet), $\chi$ and $\phi$ are DM.
We have shown that the previously found 
separation \cite{Barbieri:2006dq,LopezHonorez:2006gr,Dolle:2009fn} of the allowed
parameter space in the low- and high-mass regions for $\eta_R^0$
disappears in the presence of $\chi$ and $\phi$.

Finally, we have discussed
neutrinos coming from the annihilations of the 
captured DM  in the Sun.
The evolution equations 
of the DM numbers in the Sun,
which describe approaching
 equilibrium between the capture  and annihilation 
 (including conversion and semi-annihilation) rates of DM,
 are coupled in a multi-component DM system.
 Due to the semi-annihilations of DM,
 monochromatic neutrinos are radiated, and 
the observation rates of neutrinos have been estimated.
Observations of high-energy 
monochromatic neutrinos from the Sun may
indicate   a multi-component DM system.

\vspace*{5mm}
The work of M.~D.\ is supported by the International Max Planck Research 
School for Precision Tests of Fundamental Symmetries. 
The work of M.~A.\ is supported in part by Grant-in-Aid for Scientific 
Research for Young Scientists (B) (Grant No.22740137), 
and J.~K.\ is partially supported by Grant-in-Aid for Scientific
Research (C) from Japan Society for Promotion of Science (Grant No.22540271).
M.~D.\ thanks the Institute for Theoretical Physics at 
Kanazawa University for very kind hospitality.


\begin{thebibliography}{99}

\bibitem{Riess:1998cb}
  A.~G.~Riess {\it et al.}  [Supernova Search Team Collaboration],
``Observational Evidence from Supernovae for an Accelerating Universe and a
Cosmological Constant,''
  Astron.\ J.\  {\bf 116} (1998) 1009
  [arXiv:astro-ph/9805201].


\bibitem{Abazajian:2008wr}
  K.~N.~Abazajian {\it et al.}  [SDSS Collaboration],
``The Seventh Data Release of the Sloan Digital Sky Survey,''
  Astrophys.\ J.\ Suppl.\  {\bf 182} (2009) 543
  [arXiv:0812.0649 [astro-ph]].
  
\bibitem{Komatsu:2010fb}
  E.~Komatsu {\it et al.}  [WMAP Collaboration],
  ``Seven-Year Wilkinson Microwave Anisotropy Probe (WMAP) Observations:
 Cosmological Interpretation,''
  Astrophys.\ J.\ Suppl.\  {\bf 192} (2011) 18
  [arXiv:1001.4538 [astro-ph.CO]].

 \bibitem{Jungman:1995df}
  G.~Jungman, M.~Kamionkowski, K.~Griest,
 ``Supersymmetric dark matter,''
  Phys.\ Rept.\  {\bf 267 } (1996)  195-373
  [arXiv:hep-ph/9506380].

\bibitem{Bertone:2004pz}
  G.~Bertone, D.~Hooper and J.~Silk,
``Particle dark matter: Evidence, candidates and constraints,''
  Phys.\ Rept.\  {\bf 405} (2005) 279
  [arXiv:hep-ph/0404175];
  G.~Bertone, (ed.),
``Particle dark matter: Observations, models and searches,''
  Cambridge University Press 2010.
  
 \bibitem{Cirelli:2012tf}
  M.~Cirelli,
``Indirect Searches for Dark Matter: a status review,''
  arXiv:1202.1454 [hep-ph].
 
 
  \bibitem{Berezhiani:1990sy}
  Z.~G.~Berezhiani and M.~Y.~.Khlopov,
  ``Physics of cosmological dark matter in the theory of broken family symmetry,''
  Sov.\ J.\ Nucl.\ Phys.\  {\bf 52} (1990) 60
   [Yad.\ Fiz.\  {\bf 52} (1990) 96];
  Z.~G.~Berezhiani and M.~Y.~.Khlopov,
  ``Cosmology of Spontaneously Broken Gauge Family Symmetry,''
  Z.\ Phys.\ C {\bf 49} (1991) 73.
  
\bibitem{Boehm:2003ha}
  C.~Boehm, P.~Fayet, J.~Silk,
``Light and heavy dark matter particles,''
  Phys.\ Rev.\  {\bf D69 } (2004)  101302
  [arXiv:hep-ph/0311143].

  \bibitem{Hur:2007ur}
  T.~Hur, H.-S.~Lee and S.~Nasri,
``A Supersymmetric U(1)-prime Model with Multiple Dark Matters,''
  Phys.\ Rev.\ D {\bf 77} (2008) 015008
  [arXiv:0710.2653 [hep-ph]].

  \bibitem{Cao:2007fy}
  Q.~H.~Cao, E.~Ma, J.~Wudka and C.~P.~Yuan,
``Multipartite dark matter,''
  arXiv:0711.3881 [hep-ph].
    
  \bibitem{Feng:2008ya}
  J.~L.~Feng and J.~Kumar,
 ``The WIMPless Miracle: Dark-Matter Particles without Weak-Scale Masses or Weak Interactions,''
  Phys.\ Rev.\ Lett.\  {\bf 101} (2008) 231301
  [arXiv:0803.4196 [hep-ph]].
  
 \bibitem{SungCheon:2008ts}
  H.~S.~Cheon, S.~K.~Kang and C.~S.~Kim,
  ``Doubly Coexisting Dark Matter Candidates in an Extended Seesaw Model,''
  Phys.\ Lett.\ B {\bf 675} (2009) 203
   [Erratum-ibid.\ B {\bf 698} (2011) 324]
  [arXiv:0807.0981 [hep-ph]].
  

  \bibitem{Huh:2008vj}
  J.-H.~Huh, J.~E.~Kim, B.~Kyae,
  ``Two dark matter components in dark matter extension of the minimal supersymmetric standard model and the high energy positron spectrum in PAMELA/HEAT data,''
  Phys.\ Rev.\  {\bf D79 } (2009)  063529
  [arXiv:0809.2601 [hep-ph]].
  
  \bibitem{Fairbairn:2008fb}
  M.~Fairbairn and J.~Zupan,
  ``Dark matter with a late decaying dark partner,''
  JCAP {\bf 0907} (2009) 001
  [arXiv:0810.4147 [hep-ph]].
  
\bibitem{Zurek:2008qg}
  K.~M.~Zurek,
  ``Multi-Component Dark Matter,''
  Phys.\ Rev.\  {\bf D79 } (2009)  115002
  [arXiv:0811.4429 [hep-ph]].

  \bibitem{Morrissey:2009ur}
  D.~E.~Morrissey, D.~Poland and K.~M.~Zurek,
  ``Abelian Hidden Sectors at a GeV,''
  JHEP {\bf 0907} (2009) 050
  [arXiv:0904.2567 [hep-ph]].
 
  \bibitem{D'Eramo:2010ep}
  F.~D'Eramo and J.~Thaler,
  ``Semi-annihilation of Dark Matter,''
  JHEP {\bf 1006} (2010) 109
  [arXiv:1003.5912 [hep-ph]].

 \bibitem{Feldman:2010wy}
  D.~Feldman, Z.~Liu, P.~Nath and G.~Peim,
  ``Multicomponent Dark Matter in Supersymmetric Hidden Sector Extensions,''
  Phys.\ Rev.\  D {\bf 81} (2010) 095017
  [arXiv:1004.0649 [hep-ph]].


  \bibitem{Batell:2010bp}
  B.~Batell,
  ``Dark Discrete Gauge Symmetries,''
  Phys.\ Rev.\ D {\bf 83} (2011) 035006
  [arXiv:1007.0045 [hep-ph]].

\bibitem{Liu:2011aa}
  Z.-P.~Liu, Y.-L.~Wu and Y.-F.~Zhou,
  ``Enhancement of dark matter relic density from the late time dark matter conversions,''
  Eur.\ Phys.\ J.\ C {\bf 71} (2011) 1749
  [arXiv:1101.4148 [hep-ph]].
  
 \bibitem{Dienes:2011ja}
  K.~R.~Dienes and B.~Thomas,
  ``Dynamical Dark Matter: I. Theoretical Overview,''
  Phys.\ Rev.\ D {\bf 85} (2012) 083523
  [arXiv:1106.4546 [hep-ph]];
    K.~R.~Dienes and B.~Thomas,
  ``Dynamical Dark Matter: II. An Explicit Model,''
  Phys.\ Rev.\ D {\bf 85} (2012) 083524
  [arXiv:1107.0721 [hep-ph]];
    K.~R.~Dienes, S.~Su and B.~Thomas,
  ``Distinguishing Dynamical Dark Matter at the LHC,''
  arXiv:1204.4183 [hep-ph].
  
  
  \bibitem{Belanger:2011ww}
  G.~Belanger and J.-C.~Park,
  ``Assisted freeze-out,''
  JCAP {\bf 1203} (2012) 038
  [arXiv:1112.4491 [hep-ph]].

  \bibitem{Belanger:2012vp}
  G.~Belanger, K.~Kannike, A.~Pukhov and M.~Raidal,
  ``Impact of semi-annihilations on dark matter phenomenology - an example of $Z_N$ symmetric scalar dark matter,''
  JCAP {\bf 1204} (2012) 010
  [arXiv:1202.2962 [hep-ph]].

  \bibitem{Ivanov:2012hc}
  I.~P.~Ivanov and V.~Keus,
  ``$Z_p$ scalar dark matter from multi-Higgs-doublet models,''
  arXiv:1203.3426 [hep-ph].
  
  
  
  
 \bibitem{Ma:2006uv}
  E.~Ma,
  ``Supersymmetric Model of Radiative Seesaw Majorana Neutrino Masses,''
  Annales Fond.\ Broglie {\bf 31} (2006) 285
  [arXiv:hep-ph/0607142];
  E.~Ma,
  ``Supersymmetric U(1) Gauge Realization of the Dark Scalar Doublet Model of Radiative Neutrino Mass,''
  Mod.\ Phys.\ Lett.\  {\bf A23 } (2008)  721
  [arXiv:0801.2545 [hep-ph]].
 
     \bibitem{Fukuoka:2009cu}
  H.~Fukuoka, J.~Kubo and D.~Suematsu,
  ``Anomaly Induced Dark Matter Decay and PAMELA/ATIC Experiments,''
  Phys.\ Lett.\  B {\bf 678} (2009) 401
  [arXiv:0905.2847 [hep-ph]].

\bibitem{Fukuoka:2010kx}
  H.~Fukuoka, D.~Suematsu and T.~Toma,
  ``Signals of dark matter in a supersymmetric two dark matter model,''
  JCAP {\bf 1107} (2011) 001
  [arXiv:1012.4007 [hep-ph]].

\bibitem{Suematsu:2010nd}
  D.~Suematsu and T.~Toma,
  ``Dark matter in the supersymmetric radiative seesaw model with an anomalous
  U(1) symmetry,''
  Nucl.\ Phys.\  B {\bf 847} (2011) 567
  [arXiv:1011.2839 [hep-ph]].

\bibitem{Aoki:2011he}
  M.~Aoki, J.~Kubo, T.~Okawa and H.~Takano,
  ``Impact of Inert Higgsino Dark Matter,''
  Phys.\ Lett.\  B {\bf 707} (2012) 107
  [arXiv:1110.5403 [hep-ph]].


 


  \bibitem{Ma:2007gq}
  E.~Ma,
  ``Z(3) Dark Matter and Two-Loop Neutrino Mass,''
  Phys.\ Lett.\ B {\bf 662} (2008) 49
  [arXiv:0708.3371 [hep-ph]].

 \bibitem{Agashe:2010gt}
  K.~Agashe, D.~Kim, M.~Toharia and D.~G.~E.~Walker,
  ``Distinguishing Dark Matter Stabilization Symmetries Using Multiple Kinematic Edges and Cusps,''
  Phys.\ Rev.\ D {\bf 82} (2010) 015007
  [arXiv:1003.0899 [hep-ph]].
 

  \bibitem{Ma:2006km}
  E.~Ma,
  ``Verifiable radiative seesaw mechanism of neutrino mass and dark matter,''
  Phys.\ Rev.\  D {\bf 73} (2006) 077301
  [arXiv:hep-ph/0601225].
  
\bibitem{Griest:1988ma}
  K.~Griest,
  ``Cross-Sections, Relic Abundance and Detection Rates for Neutralino Dark
  Matter,''
  Phys.\ Rev.\  D {\bf 38} (1988) 2357
  [Erratum-ibid.\  D {\bf 39} (1989) 3802].
  
  \bibitem{Srednicki:1988ce}
  M.~Srednicki, R.~Watkins and K.~A.~Olive,
  ``Calculations of Relic Densities in the Early Universe,''
  Nucl.\ Phys.\ B {\bf 310} (1988) 693.
  
  \bibitem{Griest:1989zh}
  K.~Griest, M.~Kamionkowski and M.~S.~Turner,
  ``Supersymmetric Dark Matter Above the W Mass,''
  Phys.\ Rev.\  D {\bf 41} (1990) 3565.

  \bibitem{Kolb:1990vq}
  E.~W.~Kolb and M.~S.~Turner,
  ``The Early Universe,''
  Front.\ Phys.\  {\bf 69} (1990) 1.
  
\bibitem{Gondolo:1990dk}
  P.~Gondolo and G.~Gelmini,
  ``Cosmic abundances of stable particles: Improved analysis,''
  Nucl.\ Phys.\ B {\bf 360} (1991) 145.
    
  \bibitem{Drees:1992am}
  M.~Drees, M.~M.~Nojiri,
  ``The Neutralino relic density in minimal $N=1$ supergravity,''
  Phys.\ Rev.\  {\bf D47 } (1993)  376-408
  [arXiv:hep-ph/9207234].
  
\bibitem{Griest:1990kh}
  K.~Griest, D.~Seckel,
  ``Three exceptions in the calculation of relic abundances,''
  Phys.\ Rev.\  {\bf D43 } (1991)  3191-3203.
    
\bibitem{Ellis:1998kh}
  J.~R.~Ellis, T.~Falk and K.~A.~Olive,
  ``Neutralino - Stau coannihilation and the cosmological upper limit on the mass of the lightest supersymmetric particle,''
  Phys.\ Lett.\ B {\bf 444} (1998) 367
  [arXiv:hep-ph/9810360];
  J.~R.~Ellis, T.~Falk, K.~A.~Olive and M.~Srednicki,
  ``Calculations of neutralino-stau coannihilation channels and the cosmologically relevant region of MSSM parameter space,''
  Astropart.\ Phys.\  {\bf 13} (2000) 181
   [Erratum-ibid.\  {\bf 15} (2001) 413]
  [arXiv:hep-ph/9905481].


  
  \bibitem{Barbieri:2006dq}
  R.~Barbieri, L.~J.~Hall and V.~S.~Rychkov,
  ``Improved naturalness with a heavy Higgs: An alternative road to LHC physics,''
  Phys.\ Rev.\ D {\bf 74} (2006) 015007
  [arXiv:hep-ph/0603188].
  
\bibitem{LopezHonorez:2006gr}
  L.~Lopez Honorez, E.~Nezri, J.~F.~Oliver and M.~H.~G.~Tytgat,
``The Inert Doublet Model: An Archetype for Dark Matter,''
  JCAP {\bf 0702} (2007) 028
  [arXiv:hep-ph/0612275].

  \bibitem{Dolle:2009fn}
  E.~M.~Dolle and S.~Su,
``The Inert Dark Matter,''
  Phys.\ Rev.\ D {\bf 80} (2009) 055012
  [arXiv:0906.1609 [hep-ph]].
  

  
  \bibitem{Silk:1985ax}
  J.~Silk, K.~A.~Olive and M.~Srednicki,
  ``The Photino, the Sun and High-Energy Neutrinos,''
  Phys.\ Rev.\ Lett.\  {\bf 55} (1985) 257.

  
  \bibitem{Krauss:1985aaa}
  L.~M.~Krauss, M.~Srednicki and F.~Wilczek,
  ``Solar System Constraints and Signatures for Dark Matter Candidates,''
  Phys.\ Rev.\ D {\bf 33} (1986) 2079.
  
    
  \bibitem{Freese:1985qw}
  K.~Freese,
  ``Can Scalar Neutrinos or Massive Dirac Neutrinos Be the Missing Mass?,''
  Phys.\ Lett.\ B {\bf 167} (1986) 295.
  
  \bibitem{Gaisser:1986ha}
  T.~K.~Gaisser, G.~Steigman and S.~Tilav,
  ``Limits on Cold Dark Matter Candidates from Deep Underground Detectors,''
  Phys.\ Rev.\ D {\bf 34} (1986) 2206.
  
   \bibitem{Griest:1986yu}
  K.~Griest and D.~Seckel,
  ``Cosmic Asymmetry, Neutrinos and the Sun,''
  Nucl.\ Phys.\ B {\bf 283} (1987) 681
   [Erratum-ibid.\ B {\bf 296} (1988) 1034].
 
  \bibitem{Ritz:1987mh}
  S.~Ritz and D.~Seckel,
  ``Detailed Neutrino Spectra From Cold Dark Matter Annihilations In The Sun,''
  Nucl.\ Phys.\ B {\bf 304} (1988) 877.
  
    \bibitem{Kamionkowski:1991nj}
  M.~Kamionkowski,
  ``Energetic neutrinos from heavy neutralino annihilation in the sun,''
  Phys.\ Rev.\ D {\bf 44} (1991) 3021.

  \bibitem{Kamionkowski:1994dp}
  M.~Kamionkowski, K.~Griest, G.~Jungman and B.~Sadoulet,
  ``Model independent comparison of direct versus indirect detection of supersymmetric dark matter,''
  Phys.\ Rev.\ Lett.\  {\bf 74} (1995) 5174
  [arXiv:hep-ph/9412213].
  
  \bibitem{Cheng:2002ej}
  H.-C.~Cheng, J.~L.~Feng and K.~T.~Matchev,
  ``Kaluza-Klein dark matter,''
  Phys.\ Rev.\ Lett.\  {\bf 89} (2002) 211301
  [arXiv:hep-ph/0207125].
  
  \bibitem{Hooper:2002gs}
  D.~Hooper and G.~D.~Kribs,
  ``Probing Kaluza-Klein dark matter with neutrino telescopes,''
  Phys.\ Rev.\ D {\bf 67} (2003) 055003
  [arXiv:hep-ph/0208261].


  
    \bibitem{Ma:2001mr}
  E.~Ma and M.~Raidal,
  ``Neutrino mass, muon anomalous magnetic moment, and lepton flavor nonconservation,''
  Phys.\ Rev.\ Lett.\  {\bf 87} (2001) 011802
   [Erratum-ibid.\  {\bf 87} (2001) 159901]
  [arXiv:hep-ph/0102255].
  
      \bibitem{Hayasaka:2010et}
  K.~Hayasaka,
  ``Recent Tau Decay Results at B Factories -Lepton Flavor Violating Tau
  Decays,''
  arXiv:1010.3746 [hep-ex].

\bibitem{Nakamura:2010zzi}
  K.~Nakamura {\it et al.}  [Particle Data Group Collaboration],
  ``Review of particle physics,''
  J.\ Phys.\ G G {\bf 37} (2010) 075021
  and 2011 partial update for the 2012 edition.

  \bibitem{Gustafsson:2012aj}
  M.~Gustafsson, S.~Rydbeck, L.~Lopez-Honorez and E.~Lundstrom,
  ``Status of the Inert Doublet Model and the Role of multileptons at the LHC,''
  arXiv:1206.6316 [hep-ph].


\bibitem{Dolle:2009ft}
  E.~Dolle, X.~Miao, S.~Su and B.~Thomas,
  ``Dilepton Signals in the Inert Doublet Model,''
  Phys.\ Rev.\ D {\bf 81} (2010) 035003
  [arXiv:0909.3094 [hep-ph]];


  \bibitem{Cirelli:2005uq}
  M.~Cirelli, N.~Fornengo and A.~Strumia,
  ``Minimal dark matter,''
  Nucl.\ Phys.\ B {\bf 753} (2006) 178
  [arXiv:hep-ph/0512090].
  
  \bibitem{Schmidt:2012yg}
  D.~Schmidt, T.~Schwetz and T.~Toma,
  ``Direct Detection of Leptophilic Dark Matter in a Model with Radiative Neutrino Masses,''
  Phys.\ Rev.\ D {\bf 85} (2012) 073009
  [arXiv:1201.0906 [hep-ph]].
  
  \bibitem{Ellis:2000ds}
  J.~R.~Ellis, A.~Ferstl and K.~A.~Olive,
  ``Reevaluation of the elastic scattering of supersymmetric dark matter,''
  Phys.\ Lett.\ B {\bf 481} (2000) 304
  [arXiv:hep-ph/0001005].
  
  \bibitem{LopezHonorez:2010tb}
  L.~L. Honorez and C.~E.~Yaguna,
  ``A new viable region of the inert doublet model,''
  JCAP {\bf 1101} (2011) 002
  [arXiv:1011.1411 [hep-ph]].
  

  
   \bibitem{Aprile:2011hi}
  E.~Aprile {\it et al.}  [XENON100 Collaboration],
``Dark Matter Results from 100 Live Days of XENON100 Data,''
  Phys.\ Rev.\ Lett.\  {\bf 107} (2011) 131302
  [arXiv:1104.2549 [astro-ph.CO]].
  
   \bibitem{Angloher:2008jj}
  G.~Angloher, M.~Bauer, I.~Bavykina, A.~Bento, A.~Brown, C.~Bucci, C.~Ciemniak, C.~Coppi, G. Deuter, F. von Feilitzsch {\it et al.},
  ``Commissioning Run of the CRESST-II Dark Matter Search,''
  arXiv:0809.1829 [astro-ph].
  
  \bibitem{Lebedenko:2008gb}
  V.~N.~Lebedenko, H.~M.~Araujo, E.~J.~Barnes, A.~Bewick, R.~Cashmore, V.~Chepel, A.~Currie and D.~Davidge {\it et al.},
``Result from the First Science Run of the ZEPLIN-III Dark Matter Search Experiment,''
  Phys.\ Rev.\ D {\bf 80} (2009) 052010
  [arXiv:0812.1150 [astro-ph]].
   
\bibitem{Ahmed:2009zw}
  Z.~Ahmed {\it et al.}  [The CDMS-II Collaboration],
``Dark Matter Search Results from the CDMS II Experiment,''
  Science {\bf 327} (2010) 1619
  [arXiv:0912.3592 [astro-ph.CO]].
  
  
  \bibitem{Aalseth:2010vx}
  C.~E.~Aalseth {\it et al.}  [CoGeNT Collaboration],
``Results from a Search for Light-Mass Dark Matter with a P-type Point Contact Germanium Detector,''
  Phys.\ Rev.\ Lett.\  {\bf 106} (2011) 131301
  [arXiv:1002.4703 [astro-ph.CO]].
  
\bibitem{Bernabei:2010mq}
  R.~Bernabei {\it et al.}  [DAMA and LIBRA Collaborations],
``New results from DAMA/LIBRA,''
  Eur.\ Phys.\ J.\ C {\bf 67} (2010) 39
  [arXiv:1002.1028 [astro-ph.GA]].

\bibitem{Kim:2012rz}
  S.~C.~Kim, H.~Bhang, J.~H.~Choi, W.~G.~Kang, B.~H.~Kim, 
  H.~J.~Kim, K.~W.~Kim and S.~K.~Kim {\it et al.},
  ``New Limits on Interactions between Weakly Interacting Massive Particles and Nucleons Obtained with CsI(Tl) Crystal Detectors,''
  arXiv:1204.2646 [astro-ph.CO].


\bibitem{Bruch:2010eq}
  T.~Bruch [CDMS Collaboration],
  ``CDMS-II to SuperCDMS: WIMP search at a zeptobarn,''
  arXiv:1001.3037 [astro-ph.IM].
  
  \bibitem{Selvi:2011zz}
  M.~Selvi [XENON1T Collaboration],
  ``Study of the performances of the shield and muon veto of the XENON1T experiment,''
  PoS IDM {\bf 2010} (2011) 053.
  
  
\bibitem{Sekiya:2010bf}
  H.~Sekiya,
  ``Xmass,''
  J.\ Phys.\ Conf.\ Ser.\  {\bf 308} (2011) 012011
  [arXiv:1006.1473 [astro-ph.IM]].
  
   \bibitem{Adriani:2008zr}
  O.~Adriani {\it et al.}  [PAMELA Collaboration],
  ``An anomalous positron abundance in cosmic rays with energies 1.5-100 GeV,''
  Nature {\bf 458} (2009) 607
  [arXiv:0810.4995 [astro-ph]].

\bibitem{Barwick:1997ig}
  S.~W.~Barwick {\it et al.}  [HEAT Collaboration],
 ``Measurements of the cosmic ray positron fraction from 1-GeV to 50-GeV,''
  Astrophys.\ J.\  {\bf 482} (1997) L191
  [arXiv:astro-ph/9703192].
  \bibitem{Aguilar:2007yf}
  M.~Aguilar {\it et al.}  [AMS-01 Collaboration],
 ``Cosmic-ray positron fraction measurement from 1 to 30-GeV with AMS-01,''
  Phys.\ Lett.\ B {\bf 646} (2007) 145
  [arXiv:astro-ph/0703154].

\bibitem{FermiLAT:2011ab}
  M.~Ackermann {\it et al.}  [The Fermi LAT Collaboration],
``Measurement of separate cosmic-ray electron and positron spectra with the Fermi Large Area Telescope,''
  Phys.\ Rev.\ Lett.\  {\bf 108} (2012) 011103
  [arXiv:1109.0521 [astro-ph.HE]].
  

  \bibitem{Aharonian:2004wa}
  F.~Aharonian {\it et al.}  [HESS Collaboration],
  ``Very high-energy gamma rays from the direction of Sagittarius A*,''
  Astron.\ Astrophys.\  {\bf 425} (2004) L13
  [arXiv:astro-ph/0408145];
  F.~Aharonian {\it et al.}  [H.E.S.S. Collaboration],
  ``H.E.S.S. observations of the Galactic Center region and their possible dark matter interpretation,''
  Phys.\ Rev.\ Lett.\  {\bf 97} (2006) 221102
   [Erratum-ibid.\  {\bf 97} (2006) 249901]
  [arXiv:astro-ph/0610509].
 
    
 \bibitem{Atwood:2009ez}
  W.~B.~Atwood {\it et al.}  [LAT Collaboration],
  ``The Large Area Telescope on the Fermi Gamma-ray Space Telescope Mission,''
  Astrophys.\ J.\  {\bf 697} (2009) 1071
  [arXiv:0902.1089 [astro-ph.IM]].
  
  \bibitem{Aleksic:2009ir}
  J.~Aleksic {\it et al.}  [MAGIC Collaboration],
  ``MAGIC Gamma-Ray Telescope Observation of the Perseus Cluster of Galaxies: Implications for Cosmic Rays, Dark Matter and NGC 1275,''
  Astrophys.\ J.\  {\bf 710} (2010) 634
  [arXiv:0909.3267 [astro-ph.HE]].

  \bibitem{Acciari:2010ab}
  V.~A.~Acciari {\it et al.}  [VERITAS Collaboration],
  ``VERITAS Search for VHE Gamma-ray Emission from Dwarf Spheroidal Galaxies,''
  Astrophys.\ J.\  {\bf 720} (2010) 1174
  [arXiv:1006.5955 [astro-ph.CO]].
      
 \bibitem{Gustafsson:2007pc}
  M.~Gustafsson, E.~Lundstrom, L.~Bergstrom and J.~Edsjo,
  ``Significant Gamma Lines from Inert Higgs Dark Matter,''
  Phys.\ Rev.\ Lett.\  {\bf 99} (2007) 041301
  [arXiv:astro-ph/0703512].
  
  
  \bibitem{Suematsu:2010gv}
  D.~Suematsu, T.~Toma and T.~Yoshida,
  ``Enhancement of the annihilation of dark matter in a radiative seesaw model,''
  Phys.\ Rev.\ D {\bf 82} (2010) 013012
  [arXiv:1002.3225 [hep-ph]].

  \bibitem{Agrawal:2008xz}
  P.~Agrawal, E.~M.~Dolle and C.~A.~Krenke,
  ``Signals of Inert Doublet Dark Matter in Neutrino Telescopes,''
  Phys.\ Rev.\ D {\bf 79} (2009) 015015
  [arXiv:0811.1798 [hep-ph]].
  \bibitem{Andreas:2009hj}
  S.~Andreas, M.~H.~G.~Tytgat and Q.~Swillens,
  ``Neutrinos from Inert Doublet Dark Matter,''
  JCAP {\bf 0904} (2009) 004
  [arXiv:0901.1750 [hep-ph]].


\bibitem{IceCube:2011aj}
  R.~Abbasi {\it et al.}  [IceCube Collaboration],
``Multi-year search for dark matter annihilations in the Sun with the AMANDA-II and IceCube detectors,''
  Phys.\ Rev.\ D {\bf 85} (2012) 042002
  [arXiv:1112.1840 [astro-ph.HE]].
  
  \bibitem{Tanaka:2011uf}
  T.~Tanaka {\it et al.}  [Super-Kamiokande Collaboration],
  ``An Indirect Search for WIMPs in the Sun using 3109.6 days of upward-going muons in Super-Kamiokande,''
  Astrophys.\ J.\  {\bf 742} (2011) 78
  [arXiv:1108.3384 [astro-ph.HE]].
 
 \bibitem{Zornoza:2012tu}
  J.~D.~Zornoza,
  ``Dark matter search with the ANTARES neutrino telescope,''
  arXiv:1204.5066 [astro-ph.HE].
 
    \bibitem{Halzen:2010yj}
  F.~Halzen and S.~R.~Klein,
  ``IceCube: An Instrument for Neutrino Astronomy,''
  Rev.\ Sci.\ Instrum.\  {\bf 81} (2010) 081101
  [arXiv:1007.1247 [astro-ph.HE]].
  
  \bibitem{Scott:2012mq}
  P.~Scott {\it et al.}  [the IceCube Collaboration],
  ``Use of event-level neutrino telescope data in global fits for theories of new physics,''
  arXiv:1207.0810 [hep-ph].
  
  
 \bibitem{Esmaili:2009ks}
  A.~Esmaili and Y.~Farzan,
  ``On the Oscillation of Neutrinos Produced by the Annihilation of Dark Matter inside the Sun,''
  Phys.\ Rev.\ D {\bf 81} (2010) 113010
  [arXiv:0912.4033 [hep-ph]];
``A Novel Method to Extract Dark Matter Parameters from Neutrino Telescope Data,''
  JCAP {\bf 1104} (2011) 007
  [arXiv:1011.0500 [hep-ph]].
  
  \end{thebibliography}
\end{document}